\documentclass[iop,apj,revtex4,tighten]{emulateapj}

\usepackage{subfigure}
\def\simgt{\lower 2pt \hbox{$\, \buildrel {\scriptstyle >}\over{\scriptstyle \sim}\,$}}
\def\simlt{\lower 2pt \hbox{$\, \buildrel {\scriptstyle <}\over{\scriptstyle \sim}\,$}}

\begin{document}

\title{A comprehensive statistical assessment of star-planet
  interaction}

\author{Brendan~P.~Miller,$^{1,2,3}$ Elena~Gallo,$^{1}$
Jason~T.~Wright,$^{4,5}$ and Elliott~G.~Pearson$^{1}$}

\footnotetext[1]{Department of Astronomy, University of Michigan, Ann
  Arbor, MI 48109, USA}

\footnotetext[2]{Department of Physics and Astronomy, Macalester College,
  Saint Paul, MN 55105, USA}

\footnotetext[2]{Department of Chemistry and Physical Sciences, The
  College of St. Scholastica, Duluth, MN 55811, USA}

\footnotetext[4]{Department of Astronomy and Astrophysics, The
  Pennsylvania State University, University Park, PA 16802, USA}

\footnotetext[5]{Center for Exoplanets and Habitable Worlds, The
  Pennsylvania State University, University Park, PA 16802}

\begin{abstract}

We investigate whether magnetic interaction between close-in giant
planets and their host stars produce observable statistical
enhancements in stellar coronal or chromospheric activity. New {\it
  Chandra\/} observations of 12 nearby ($d<60$~pc) planet-hosting
solar analogs are combined with archival {\it Chandra\/}, {\it
  XMM-Newton\/}, and {\it ROSAT\/} coverage of 11 similar stars to
construct a sample inoculated against inherent stellar class and
planet-detection biases. Survival analysis and Bayesian regression
methods (incorporating both measurements errors and X-ray upper
limits; 13/23 stars have secure detections) are used to test whether
``hot Jupiter'' hosts are systematically more X-ray luminous than
comparable stars with more distant or smaller planets.  No significant
correlations are present between common proxies for interaction
strength ($M_{\rm P}/a^{2}$ or $1/a$) versus coronal activity ($L_{\rm
  X}$ or $L_{\rm X}/L_{\rm bol}$). In contrast, a sample of 198 FGK
main-sequence stars does show a significant ($\sim99$\% confidence)
increase in \hbox{X-ray} luminosity with $M_{\rm P}/a^{2}$. While
selection biases are incontrovertibly present within the main-sequence
sample, we demonstrate that the effect is primarily driven by a
handful of extreme hot-Jupiter systems with $M_{\rm
  P}/a^{2}>450$~$M_{\rm Jup}$~AU$^{-2}$, which here are all X-ray
luminous but to a degree commensurate with their \ion{Ca}{2} H and K
activity, in contrast to presented magnetic star-planet interaction
scenarios that predict enhancements relatively larger in $L_{\rm
  X}$. We discuss these results in the context of cumulative tidal
spin-up of stars hosting close-in gas giants (potentially followed by
planetary infall and destruction). We also test our main-sequence
sample for correlations between planetary properties and UV luminosity
or \ion{Ca}{2} H and K emission, and find no significant dependence.

\end{abstract}


\section{Introduction}

The initial detection of ``hot Jupiter'' exoplanets (with $M_{\rm
  P}\simgt1 M_{\rm Jup}$ but in orbits with semi-major axes only a
fraction that of Mercury) presented a challenge to traditional models
of planetary formation. While it is now clear that such systems are
exceptional rather than the norm (e.g., Wright et al.~2012), and
likely result from planet-disk or planet-planet migration, these
extreme cases permit investigation of planetary events that are
irrelevant or observationally inaccessible for longer-period
planets. For example, close-in gas giants\footnote{Or smaller planets
  as well (Rappaport et al.~2012; Budaj 2013).} can experience
substantial mass loss due to atmospheric heating and inflation from
high-energy X-ray and far ultraviolet (FUV) irradiation (Lammer et
al.~2003; Knutson et al.~2010), and tidal effects may also help strip
planetary atmospheres (Jackson et al.~2010). X-ray observations are
just beginning to constrain such theories (e.g., Poppenhaeger et
al.~2012). Tidal decay may destroy hot Jupiters on Gyr timescales, as
suggestively supported by an apparent scarcity of close-in massive
planets around older stars (Jackson et al.~2009; Debes \& Jackson
2010). Such studies also explore how hot Jupiters might spin-up their
host stars, and help put constraints on tidal $Q$ values for stars and
giant exoplanets. Magnetic star-planet interaction could speed
evaporation of planetary atmospheres (e.g., Lanza 2013) and may
enhance stellar coronal and chromospheric activity in hot Jupiter
systems (Cuntz et al.~2000; Rubenstein \& Schaefer 2000), to a degree
proportional to the exoplanet magnetic field strength. Here we
investigate the statistical observability of star-planet interaction,
primarily by means of single-pointing \hbox{X-ray} measurements
tracing coronal activity.

The energy released in a magnetic star-planet interaction event is
theoretically expected to scale approximately as $B_{\rm *}B_{\rm
  P}v_{\rm rel}a^{-n}$ (Saar et al.~2004; Kashyap et al.~2008;
alternatively the radius of the planetary magnetosphere may be
parameterized in terms of $B_{\rm *}$ and $B_{\rm P}$ and then the
dissipated power scales as $B_{\rm *}^{4/3}B_{\rm P}^{2/3}$, as in
Lanza~2009; see also Cuntz et al.~2000). Here $B_{\rm *}$ and $B_{\rm
  P}$ are the star and planet magnetic field strengths, respectively,
and $v_{\rm rel}$ is the relative velocity between field
lines.\footnote{The relative velocity in terms of observables is
  $K(R_{\rm *}/a)-v_{\rm rot}$ with $v_{\rm rot}=v\sin{i_{\rm
      *}}/\sin{i_{\rm *}}$ at the stellar equator (Cuntz et
  al.~2000).}  The dependence upon the semi-major axis $a$ is $n\sim3$
close to the star (for a simple dipole) and $n\sim2$ further out, in
the ``Parker spiral'' region. The stellar magnetic field strength can
be measured directly (e.g., Fares et al.~2010 find averages of 33, 22,
and 36 Gauss in 2006, 2007, and 2008 for HD~189733, and Fares et
al.~2012 find averages of 2.6 and 3.7 Gauss in 2007 and 2009 for
HD~179949; note the actual field structure is not a simple dipole but
is complex in both these hot Jupiter hosts) or estimated from the
intrinsic X-ray-to-bolometric luminosity $L_{\rm X}/L_{\rm bol}$ (in
the absense of any planetary enhancement; Pevtsov et al.~2003), which
in turn scales with the chromospheric activity measurable in
\ion{Ca}{2} H and K line core emission and parameterized by $R_{\rm
  HK}^{'}$ (Mamajek \& Hillenbrand 2008).

The planetary magnetic field strength is unknown; it is potentially
measureable with a calibrated star-planet interaction relation, or
speculatively via radio emission (Grie{\ss}meier et al.~2007; Fares et
al.~2010; Lecavelier des Etangs et al.~2013) or bow-shock produced
offsets in UV transit times (Vidotto et al.~2011; but see also Turner
et al.~2013, 2014). Based on direct measurements within our own solar
system, planetary magnetic field strength likely scales with mass
(Arge et al~1995; Stevens 2005); it may also depend upon rotation
rate, which for close-in gas giants is tidally locked to the orbital
period (Bodenheimer et al.~2001) of days, rather than the $\sim$10
hours for Jupiter (which, for reference, has an equatorial field
strength of order 4.3 gauss). The best radial-velocity selected
candidates to display star-planet interaction based on the relation
$B_{\rm *}B_{\rm P}v_{\rm rel}a^{-n}$ include well-studied hot Jupiter
systems such as $\upsilon$~And, $\tau$~Boo, HD 75289, HD 179949, HD
189733, and HD 209458, all of which rank in the top 10\% of predicted
energy released. For a fixed set of stellar parameters, the
interaction energy is expected to scale simply with $M_{\rm P}/a^{2}$
(the systems listed above are also in the top 10\% by this metric),
and we use this proxy\footnote{We ignore the distinction between
  $M_{\rm P}$ and the minimum mass of explanets derived from radial
  velocity work, since the $\sin{i}$ inclination term is unknown for
  most of these systems but provides a typical correction of only
  $\sim$15\% (Wright \& Gaudi 2013).} throughout (see also Miller et
al.~2012), along with $1/a$ for comparison to other studies.

Some numerical work supports the theory of magnetic star-planet
interaction. Lanza~(2008) modeled chromospheric hot spots in several
systems (offset from the subplanetary point by varying degrees) as
arising from star-planet magnetic reconnection events. Saur~(2013)
pointed out that only limited energy fluxes are expected from
sub-Alfvenic plasma interactions, however Lanza~(2009) suggested that
interaction may serve as a catalyst for a release of coronal field
energy. Cohen et al.~(2009, 2011) carried out three-dimensional
magnetohydrodynamic simulations illustrating that close-in giant
planets can potentially produce an increase in overall \hbox{X-ray}
luminosity, and generate (non-persistent) coronal hot spots that
rotate synchronously with the planet (albeit potentially shifted in
phase); see also Pillitteri et al.~(2010). Observational evidence of
magnetic star-planet interaction has now been claimed for several
individual cases (including five of the six systems mentioned above,
excepting HD~209458; e.g., Shkolnik et al.~2005; Saar et al.~2008;
Walker et al.~2008; Pillitteri et al.~2010; Lanza et al.~2011) based
on measurements of photospheric (optical light curve), chromospheric
(\ion{Ca}{2} H and K line core emission) or coronal (\hbox{X-ray}
emission) activity enhancements concentrated near a specific planetary
orbital phase. However, even in the best candidate systems, any
signatures seem to be transient (Shkolnik et al.~2008). As a striking
counterexample, the extreme WASP-18 system, which contains a $\simeq10
M_{\rm Jup}$ planet in a 0.94~d period, shows no evidence for
planet-linked activity (Miller et al.~2012; Pillitteri et al.~2014b);
such interaction must be either absent, only rarely present, or
rendered inefficient by an extremely weak stellar magnetic
field\footnote{WASP-18 has $R_{\rm HK}^{'}=-5.15$ (Miller et
  al.~2012), which is in the bottom $\sim$10\% for activity of stars
  known to host hot Jupiters.} (Lanza et al.~2013; Shkolnik et
al.~2013; but see also Miller et al.~2012).

Single-pointing \hbox{X-ray} observations of planet-bearing stars have
been utilized to search for systematic enhancements in $L_{\rm X}$ in
hot Jupiter systems, with mixed results (e.g., Kashyap et al.~2008;
Poppenhaeger et al.~2010; Scharf 2010; Poppenhaeger \& Schmitt
2011). Such investigations typically statistically average over
orbital phase, trading decreased sensitivity to a phase-restricted
effect in a given system for a much larger sample size and increased
comparative sensitivity to phase-independent or full-surface activity
increases; these \hbox{X-ray} surveys probe magnetic or tidal
star-planet interaction (presumed one-sided or two-sided,
respectively) as well as the potential cumulative tidal influence of
hot Jupiters upon their hosts (e.g., stellar spin-up and associated
activity rejuvenation; Schr{\"o}ter et al.~2011; Poppenhaeger et
al.~2013). However, this approach is challenged by deeply embedded
biases. For example, it is more difficult with radial velocity (RV)
searches to detect distant or low-mass planets around intrinsically
active stars, so in an RV-derived sample the hot Jupiter systems will
have relatively greater average X-ray luminosities already prior to
considering any star-planet interaction. While Kashyap et al.~(2008)
found that stars hosting planets with $a<0.15$~AU remain X-ray
brighter by a factor of 1.3--4 even after attempting to control for
this sensitivity bias, Poppenhaeger et al.~(2010) found no significant
trends in $L_{\rm X}/L_{\rm bol}$ with $M_{\rm P}$ or $a$ and ascribed
a weak observed correlation between $L_{\rm X}$ with $M_{\rm P}/a$
entirely to such selection effects. Other biases may result from
inhomogeneous stellar properties and/or Malmquist-type distance
incompleteness, particularly with shallow {\it ROSAT\/} All-Sky Survey
(RASS) data. A RASS-based study by Scharf (2010) did not support
comparatively higher stellar $L_{\rm X}$ in systems with $a<0.15$~AU
planets, but did note a strong correlation between $L_{\rm X}$ and
$M_{\rm P}$ in such short-period systems; however, Poppenhaeger \&
Schmitt (2011) demonstrated that this trend is not significantly
present in $L_{\rm X}/L_{\rm bol}$ with deeper {\it XMM-Newton\/}
data.

\begin{figure}
\includegraphics[scale=0.45]{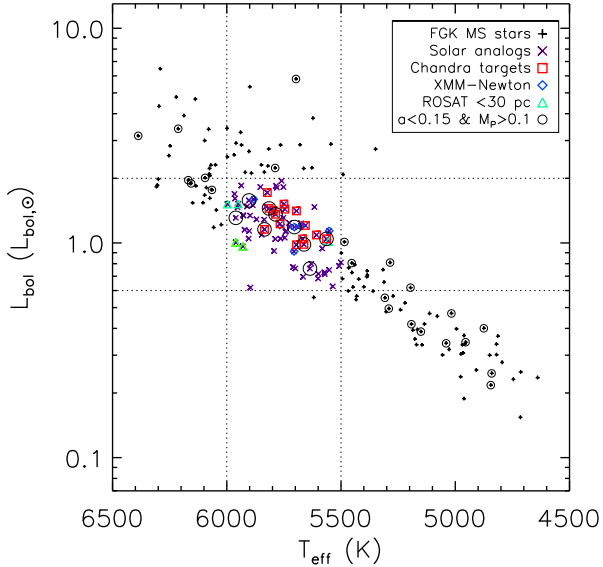} \figcaption{\small HR
  diagram for the full sample of FGK main-sequence planet-hosting
  stars. The dashed lines mark the $5500<T_{\rm eff}<6000$ and
  $0.6<L_{\rm bol}/L_{\odot}<2$ boundaries that we use to define solar
  analogs; coverage by {\it Chandra\/} ($d<60$~pc), {\it XMM-Newton\/}
  ($d<30$~pc), and {\it ROSAT\/} ($d<30$~pc) is indicated with red
  squares, blue diamonds, and green triangles, respectively. Stars
  hosting close-in, massive planets ($a<0.15$ AU and $M_{\rm P}>0.1
  M_{\rm Jup}$) are also circled.}
\end{figure}

We are able to improve upon previous studies in several areas. Most
obviously, many planetary systems have been newly discovered in the
last few years. This gives a larger pool of candidates from which to
draw for statistical study, which in turn permits us to be more
discriminating than was previously practical. We only use published
and verified planets discovered in radial velocity searches. Our
main-sequence (MS) sample carefully excludes stars for which the color
and bolometric luminosity indicate even modest post-MS
evolution. Further, we incorporate recent {\it XMM-Newton\/} (e.g.,
Poppenhaeger et al.~2010) and {\it Chandra\/} (e.g., this work)
observations that detect stars down to luminosities far below shallow
{\it ROSAT\/} limits, which provides a larger detection fraction that
improves both the sensitivity and reliability of correlation or
regression tests. In contrast to most previous statistical studies, we
consider orbital phase where available. Complementary multi-wavelength
data is now published for most systems, and we additionally
investigate UV luminosity and \ion{Ca}{2} H and K emission, and
consider our \hbox{X-ray} results in the context of the intrinsic
chromospheric activity. The focus of this work is a study of solar
analogs, with the sample constructed to mitigate many of the selection
biases that affected previous work (e.g., Poppenhaeger \& Schmitt
2011). Finally, we use a variety of statistical tests to assess
whether hot Jupiter systems display enhanced activity, including the
Bayesian linear regression tool of Kelly (2007) that handles both
measurement errors and censoring. This last step of fully accounting
for \hbox{X-ray} upper limits is critical to obtaining an accurate
determination of significance for any potential correlations. Our work
here provides the most comprehensive statistical assessment of the
observability of star-planet interaction conducted to date.

This paper is organized as follows. Section 2 describes the
construction of the main-sequence and solar analogs samples and the
X-ray observations and data reduction. Section 3 presents the results
of testing the solar analogs and full samples for enhanced X-ray
emission in hot Jupiter systems, including consideration of selection
biases. Section 4 considers several interpretations of these
results. Section 5 investigates other measures of stellar activity,
specifically UV luminosity and \ion{Ca}{2} H and K emission. Section 6
summarizes and provides our conclusions.

\begin{figure}
\includegraphics[scale=0.45]{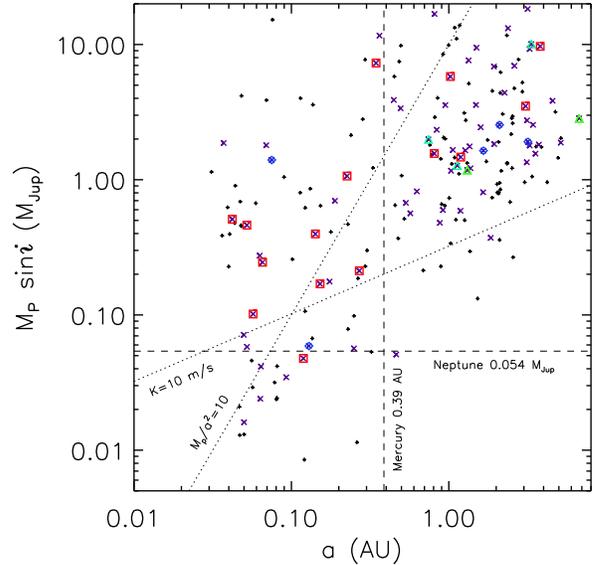} \figcaption{\small
  Semi-major axis versus minimum planetary mass for the most strongly
  interacting planet in each stellar system (symbols as in
  Figure~1). The semi-major axis of Mercury and the mass of Neptune
  are indicated for reference (with vertical and horizontal dotted
  lines, respectively). The marked dashed lines show $K=10 m s^{-1}$
  (for a solar-mass star) and $\log{M_{\rm P}/a^{2}}=10 M_{\rm
    Jup}$~AU$^{-2}$, which illustrate radial velocity completeness and
  our preferred proxy for interaction strength, respectively, on
  semi-major axis and planetary mass.}
\end{figure}

\begin{deluxetable*}{p{50pt}rrrrrrrrrrrrr}
\tablecaption{Sample properties}
\tabletypesize{\scriptsize}
\tablewidth{18.5cm}

\tablehead{\colhead{Name} & \colhead{$T_{\rm eff}$} & \colhead{$L_{\rm
  bol}$} & \colhead{$R_{\rm HK}^{'}$} & \colhead{RMS} &
  \colhead{$M\sin{i}$} & \colhead{$a$} & \colhead{Per} & \colhead{$K$}
  & \colhead{$M_{\rm P}/a^{2}$} & \colhead{$1/a$} & \colhead{$L_{\rm
  X}$} & \colhead{$\frac{L_{\rm X}}{L_{\rm bol}}$} & \colhead{Obs\tablenotemark{a}} \\
  \colhead{} & \colhead{(K)} & \colhead{(erg s$^{1}$)} & \colhead{} &
  \colhead{(m s$^{-1}$)} & \colhead{($M_{\rm Jup}$)} & \colhead{(AU)} &
  \colhead{(d)} & \colhead{(m s$^{-1}$)} & \colhead{($\frac{M_{\rm Jup}}{
  AU^{2}}$)} & \colhead{(AU$^{-1}$)} & \colhead{(erg s$^{-1}$)} &
  \colhead{} & \colhead{}}

\startdata
HD 142 b &         6248 &      34.04 &      -4.92 &      12.00 &       1.31 &       1.04 &     350.30 &      33.90 &       1.20 &       0.96 &   $<$27.90 &          $<-$6.14 &   \\ 
HD 1237 b &        5536 &      33.38 &      -4.44 &      19.20 &       3.37 &       0.49 &     133.71 &     167.00 &      13.79 &       2.02 &      28.75$\pm$0.16 &  $-$4.63 & B \\ 
HD 1461 b &        5765 &      33.60 &      -5.02 &       3.43 &       0.02 &       0.06 &       5.77 &       2.70 &       5.95 &      15.74 &   $<$27.81 &          $<-$5.79 &   \\ 
HIP 2247 b &       4714 &      32.98 &      -4.79 &       4.10 &       5.12 &       1.34 &     655.60 &     173.30 &       2.86 &       0.75 &   $<$28.18 &          $<-$4.80 &   \\ 
HD 2638 b &        5192 &      33.21 &     -99.00 &       3.30 &       0.48 &       0.04 &       3.44 &      67.40 &     251.39 &      22.95 &   $<$28.47 &          $<-$4.74 &   \\ 
HD 3651 b &        5220 &      33.31 &      -5.02 &       6.30 &       0.23 &       0.29 &      62.22 &      15.90 &       2.64 &       3.39 &      27.12$\pm$0.18 &  $-$6.19 & H \\ 
HD 4113 b &        5688 &      33.66 &     -99.00 &       8.40 &       1.65 &       1.27 &     526.62 &      97.10 &       1.02 &       0.79 &   $<$28.37 &          $<-$5.29 &   \\ 
HD 4208 b &        5600 &      33.42 &      -4.95 &       3.40 &       0.81 &       1.65 &     828.00 &      19.06 &       0.30 &       0.60 &   $<$28.10 &          $<-$5.32 &   \\ 
{\bf HD 4308 b} &  5695 &      33.58 &     -99.00 &       1.30 &       0.05 &       0.12 &      15.56 &       4.07 &       3.36 &       8.39 &      26.66$\pm$0.14 &  $-$6.92 & C \\ 
HD 5388 b &        6297 &      34.22 &      -4.98 &       3.33 &       1.97 &       1.76 &     777.00 &      41.70 &       0.63 &       0.57 &   $<$28.54 &          $<-$5.68 &   \\ 
{\bf HD 6434 b} &  5835 &      33.65 &      -4.90 &      10.60 &       0.40 &       0.14 &      22.00 &      34.20 &      19.68 &       7.04 &   $<$26.60 &          $<-$7.05 & C \\ 
HIP 5158 b &       4962 &      32.86 &      -4.80 &       2.47 &       1.43 &       0.89 &     345.72 &      57.00 &       1.81 &       1.13 &   $<$28.31 &          $<-$4.55 &   \\ 
HD 6718 b &        5746 &      33.61 &      -4.97 &       1.79 &       1.56 &       3.55 &    2496.00 &      24.10 &       0.12 &       0.28 &   $<$28.56 &          $<-$5.05 &   \\ 
HD 7199 b &        5386 &      33.46 &      -4.80 &       2.63 &       0.30 &       1.36 &     615.00 &       7.76 &       0.16 &       0.73 &   $<$28.19 &          $<-$5.27 &   \\ 
HD 7449 b &        6024 &      33.67 &      -4.85 &       3.81 &       1.31 &       2.34 &    1275.00 &      41.59 &       0.24 &       0.43 &   $<$28.25 &          $<-$5.42 &   \\ 
HD 7924 b &        5177 &      33.14 &      -4.89 &       2.78 &       0.03 &       0.06 &       5.40 &       3.87 &       9.08 &      17.66 &      27.27$\pm$0.26 &  $-$5.87 & F \\ 
HD 8535 b &        6136 &      33.85 &      -4.95 &       2.49 &       0.68 &       2.44 &    1313.00 &      11.80 &       0.11 &       0.41 &   $<$28.52 &          $<-$5.33 &   \\ 
HD 8574 b &        6049 &      33.95 &      -5.09 &      14.20 &       1.81 &       0.76 &     227.00 &      58.30 &       3.15 &       1.32 &   $<$28.38 &          $<-$5.57 &   \\ 
HD 9446 b &        5793 &      33.55 &      -4.50 &      15.10 &       0.70 &       0.19 &      30.05 &      46.60 &      19.52 &       5.29 &      28.49$\pm$0.28 &  $-$5.06 & F \\ 
$\upsilon$ And b &    6212 &      34.12 &      -4.98 &      13.66 &       0.67 &       0.06 &       4.62 &      68.21 &     189.78 &      16.84 &      27.66$\pm$0.12 &  $-$6.46 & O \\[-15pt]
\enddata

\tablecomments{Table is ordered by RA. The 23 bolded names make up the
  solar analogs subsample (the two italicized names are excluded due
  to high $R_{\rm HK}^{'}$ values). The full table is available
  online; a portion is provided here to illustrate format.}

\tablenotetext{a}{Observatory: no entry is RASS limit; B/F is RASS
  Bright/Faint source; P/H is {\it ROSAT\/} pointed PSPC/HRI
  detection; X is {\it XMM-Newton\/} detection or limit; C is {\it
    Chandra\/} detection or limit; O is other as explained in text.}

\end{deluxetable*}

\section{Sample construction and X-ray luminosity measurements}

\subsection{Main sequence FGK and solar analogs selection}

We selected main-sequence planet-hosting stars from the Exoplanet
Orbit Database (EOD)\footnote{{\tt http://exoplanets.org}} as of March
2013. Our sample is limited to stars at distances less than 60
parsecs, and restricted to RV discovered planets with $M\sin{i}<20
M_{\rm Jup}$ (this is deliberately greater than the usual brown dwarf
cutoff so as to include all strongly interacting systems, but only HD
162020 has $M\sin{i}>13 M_{\rm Jup}$ and $a<0.15$~AU). Stars are
identified as main sequence by requiring surface gravity $\log{g}>3.8$
and inferred stellar radius $R_{\rm *}<2 R_{\odot}$, which is
appropriate for non-evolved stars across the full range of
temperatures here considered. (See, e.g., Niedzielski et al.~2014 and
references therein for planet-hosting evolved systems.) We require the
effective temperature to be between 4000~K and 6500~K. The upper
temperature limit excludes early F and hotter stars for which the
disappearance of the upper convective zone eliminates dynamo action
and leaves them generally X-ray dark. The lower temperature limit
excludes M dwarfs, which have efficient dynamos (Kitchatinov \&
Olemskoy 2011) and coronal activity producing high $L_{\rm X}/L_{\rm
  bol}$ ratios (they are also generally UV active; France et
al.~2013). M dwarfs are also less likely to host gas giants and more
likely to contain super Earths (e.g., Wu \& Lithwick 2013); this is
not solely a selection effect, as theoretical models (e.g., Boss 2006)
predict that low-mass stars more easily (but certainly not
exclusively; see also Johnson et al.~2012) form lower-mass planets.

For multiple-planet systems, we consider only the properties of the
planet most relevant for potential magnetic interaction. We rank by
$M_{\rm P}/a^{2}$ in multi-planet systems (see discussion in Section
1); in nearly all cases this is equivalent to choosing the closest-in
known planet. The bolometric luminosity is calculated from the B-V
color index and the apparent V-band magnitude $m_{\rm v}$ based on the
main-sequence trends tabulated by Bessell et al.~(1998); specifically,
we take the bolometric correction to be $0.59 - 2.32(B-V) +
2.60(B-V)^{2} -0.48(B-V)^{3}$. Measurements of $R_{\rm HK}^{'}$ are
from the EOD except for five values for hot Jupiter systems adopted
from Knutson et al.~(2010)\footnote{Use of this catalog instead of the
  values in the EOD changed HD 179949 from $-$4.8046 to $-$4.622, HD
  189733 from no measurement to $-$4.501, HD 209458 from $-$5.014 to
  $-$4.970, HD 80606 from $-$5.0886 to $-$5.061, and $\upsilon$~And
  from $-$5.066 to $-$4.982.}  and eight values added from Isaacson \&
Fischer (2010).\footnote{Specifically, HD 28185, 16760, 102365, 37603,
  114762, 155358, 156846, and 171238 have $R_{\rm HK}^{'}$ values of
  $-$5.023, $-$4.923, $-$4.931, $-$5.025, $-$4.902, $-$4.931,
  $-$5.082, $-$4.605.}  Measurements of $L_{\rm UV}/L_{\rm bol}$ are
taken from Shkolnik (2013).

From the main-sequence parent sample of planet-hosting stars, we
select solar analogs as having $5500 K < T_{\rm eff} < 6000 K$ and
$0.6 < L_{\rm bol}/L_{\odot} < 2$. Additional criteria were imposed on
our new {\it Chandra\/} targets: those 12 stars are non-active
($v\sin{i} < 2$ km s$^{-1}$ and $R^{'}_{\rm HK}<-4.8$ where known) and
have definitively Jovian planets ($M_{\rm P}>0.1 M_{\rm Jup}$) in low
eccentricity orbits ($e<0.3$). The activity and mass cuts further
reduce detectability bias, and additionally sub-Jovian planets may
have distinct magnetic field properties that could blur a statistical
signature of interaction in the sample. Low eccentricity establishes
that the planet remains at a nearly constant distance from its star,
so any interaction is quasi-continuous and \hbox{X-ray} observations
do not need to be timed to periastron (but see also Hodgson et
al.~2014 for using eccentric systems to test for star-planet
interaction).

Targeting low-activity solar analogs reduces detectability bias, but
it has both additional advantages and disadvantages for investigating
star-planet interaction. The absolute energy produced by
planet-induced activity is expected to scale with stellar magnetic
field strength ($\S$1); however, the fractional increase in
\hbox{X-ray} luminosity may be relatively constant, due to $L_{\rm X}$
correlating with $B_{\rm *}$ (Pevtsov et al.~2003). In active systems
the intrinsic chromospheric or coronal variability, such as from
starspots or flares, is generally several times greater than that
expected from star-planet interaction (e.g., see discussion of
HD~73256 in Shkolnik et al.~2005). Intensive phase-resolved coverage
of individual active systems can potentially distinguish stellar from
planet-induced variability (Shkolnik et al.~2005, 2008; Miller
al.~2012; Scandariato et al.~2013), but for our statistical study that
seeks to test planet-induced enhancements in \hbox{X-ray} luminosity
by an average factor of a few (Kashyap et al.~2008) it is preferable
to mitigate against increased noise. After considering the properties
of solar analogs, we broaden the scope of our analysis to the full FGK
MS sample that also includes more active stars.

Properties of the 198 FGK MS stars and their relevant planets are
provided in Table~1. Figure~1 shows an HR diagram of the full sample,
with the cuts selecting solar analogs indicated as dashed
lines. Figure~2 shows the semi-major axis and planetary mass for the
most relevant planet in each system, as defined above; there is good
coverage of the $a-M_{\rm P}$ plane. Trend lines for constant
semi-amplitude $K{\propto}M_{\rm P}/a^{0.5}$ (which, along with
intrinsic stellar noise, determines detection sensitivity) and for
expected interaction strength $M_{\rm P}/a^{2}$ are also shown on
Figure~2.

\subsection{Chandra and archival X-ray observations}

{\it Chandra\/} archival coverage of solar analogs includes
observations of 51 Peg (ObsID 10825; PI Schmitt; Poppenhaeger et
al.~2009), HD 4308 (ObsID 12339; PI Schmitt), and $\rho$ CrB (ObsID
12396; PI Saar; Saar \& Testa 2012). We reprocessed these data as
described below. {\it XMM-Newton\/} observations targeting
planet-hosting stars within $d<30$~pc (Kashyap et al.~2008;
Poppenhaeger et al.~2010) include an five additional solar analogs (47
UMa, HD 190360, HD 217107, 16 Cyg B, and HD 70642; 51 Peg and HD 4308
also have {\it XMM-Newton\/} coverage), for which we take $L_{\rm X}$
measurements from Poppenhaeger et al.~(2010). For completeness we also
include complementary {\it ROSAT\/} coverage of five solar analogs at
$d<30$~pc\footnote{We do not include the shallow {\it ROSAT\/}
  coverage at $30<d<60$~pc, which provides only a $\sim$6\% detection
  rate, in our analysis of the solar analogs, but it is used for the
  full FGK MS sample.}
(HD 39091, HD 82943, HD 147513, HD 150706, HD 210277), calculating
$L_{\rm X}$ from RASS bright or faint source catalog net count rates.
However, two (HD 147513 and HD 150706) of these five solar analogs
with {\it ROSAT\/} coverage are quite active (likely because they are
young), with $\log{R^{'}_{\rm HK}}>-4.5$, and they are set aside for
this analysis as they do not provide a proper point of comparison with
true solar analogs; note that they both have only distant Jupiter-mass
planets known, so their activity cannot be related to star-planet
interaction. To this archival data we add our 12 new {\it Chandra\/}
observations (PI Miller, ObsIDs 13658--13669) to construct a sample of
23 solar analogs with sensitive \hbox{X-ray} coverage.

The new and archival {\it Chandra\/} observations of solar analogs are
detailed in Table~2. These observations were taken with the ACIS-S
array, with the target positioned at the aim point of the S3 chip; the
soft-band sensitivity of this back-illuminated chip is helpful for
detecting coronal emission (Poppenhaeger et al.~2009). All
observations were taken using Very Faint telemetry to optimize
background removal. The data were reduced using the CIAO software
package, version 4.5, using standard techniques which are briefly
described below.

The data were reprocessed with the EDSER subpixel optimization
applied, with Very Faint particle background cleaning applied, with
the charge transfer inefficiency correction applied, with
time-dependent gain correction applied, and using the most recently
available CALDB calibration files (including the updated ACIS
contamination model that accounts for condensation on the optical
blocking filters). The standard grades of 02346 were
retained. Observations were checked for background flaring; point
sources were identified on the S3 chip using {\it wavdetect\/} and
removed for these purposes, then the {\it deflare\/} script was run on
a lightcurve binned to 200~s and any identified intervals of high
background were filtered out of the level 2 event file. A 0.8~keV
exposure map was used to determine the effective exposure times at the
source positions, and to excise edge regions ($<$1.5\%). A 0.15--2~keV
image of the S3 chip was created for each observation and used for all
subsequent analysis. The coronal emission from these solar-type stars
is not expected to extend to harder energies (we verified that at most
a few percent of the source counts have energies $>2$~keV), and
insufficient counts are available to justify subdividing this energy
band.

\begin{figure*}
\includegraphics[scale=0.95]{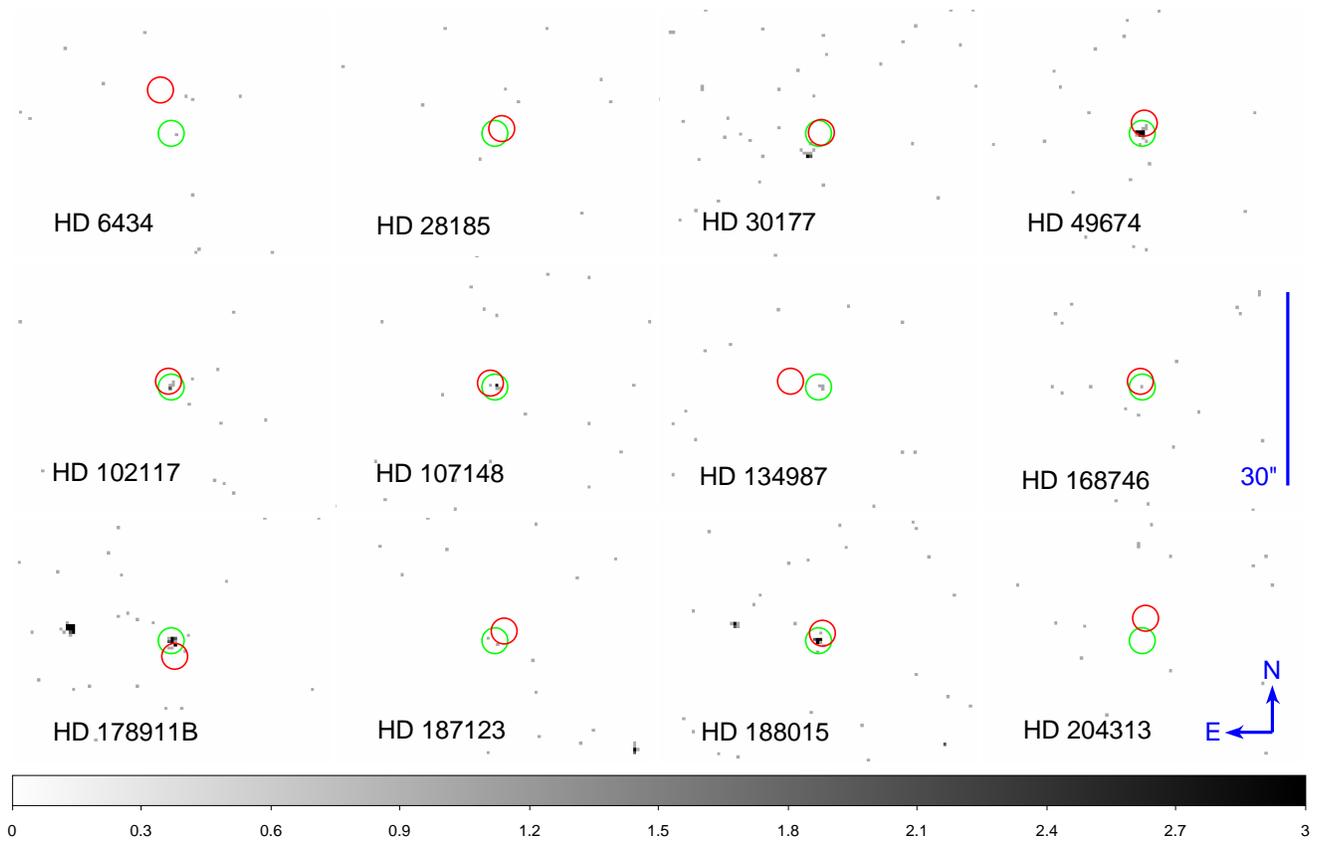} \figcaption{\small
  {\it Chandra\/} 0.2--2 keV images of our 12 newly observed solar
  analogs. The cutouts are ordered by RA, marked by name, and plotted
  with a fixed linear grayscale. The red and green circles are $r=2''$
  apertures placed at the J2000 and observation epoch for each star,
  calculated from {\it Hipparcos\/} proper motions. The exposures and
  count rates are given in Table~2.}
\end{figure*}

\begin{deluxetable*}{p{47pt}rrrrrrrrrrrr}
\tablecaption{Chandra observations of solar analogs}
\tabletypesize{\scriptsize}
\tablewidth{17.3cm}
\tablehead{\colhead{Name} & \colhead{ObsID} & \colhead{HJD} &
  \colhead{$\phi$} & \colhead{Exp} & \colhead{Cts} & \colhead{Rate} &
  \colhead{Flux} & \colhead{Dist} & \colhead{$L_{\rm X}$} &
  \colhead{$m_{\rm v}$} & \colhead{$L_{\rm bol}$} &
  \colhead{$\frac{L_{\rm X}}{L_{\rm bol}}$}}

\startdata
HD 4308    & 12339\tablenotemark{a}	& 2455622.92    & 0.097   & 14.88  &    20.68      &    1.39    &    5.49   & 22.06    &    26.66$\pm$0.14     & 6.55 & 33.58    &    $-$6.92 \\
HD 6434    & 13662	& 2456202.55	& 0.430   &  8.97  & $<$3.12       & $<$0.35    & $<$1.38   & 41.37    & $<$26.60              & 7.72 & 33.65    & $<$$-$7.05 \\
HD 28185   & 13664	& 2456202.67	& 0.092   &  8.68  & $<$3.17       & $<$0.37    & $<$1.46   & 42.34    & $<$26.64              & 7.80 & 33.67    & $<$$-$7.03 \\
HD 30177   & 13668	& 2456117.61	& 0.483   & 14.88  & $<$3.31       & $<$0.22    & $<$0.87   & 52.83    & $<$26.62              & 8.41 & 33.62    & $<$$-$7.00 \\
HD 49674   & 13661	& 2456263.06	& 0.960   & 11.94  &    42.71      &    3.58    &    14.13  & 44.23    &    27.67$\pm$0.11     & 8.10 & 33.58    &    $-$5.91 \\
HD 102117  & 13663	& 2456196.61	& 0.991   &  7.00  &    4.93       &    0.70    &    2.76   & 39.70    &    26.87$\pm$0.27     & 7.47 & 33.73    &    $-$6.86 \\
HD 107148  & 13665	& 2456248.46	& 0.704   &  9.57  &    6.91       &    0.72    &    2.84   & 51.20    &    27.10$\pm$0.22     & 8.01 & 33.74    &    $-$6.64 \\
HD 134987  & 13666	& 2455937.42	& 0.385   &  4.99  &    2.92\tablenotemark{b}       &    0.58    &    2.29   & 26.21    &    26.43$\pm$0.39     & 6.47 & 33.77    &    $-$7.34 \\
$\rho$ CrB    & 12396\tablenotemark{a}	& 2455944.05	& 0.644   &  9.86  &    5.88       &    0.60    &    2.37   & 17.24    &    26.07$\pm$0.24     & 5.39 & 33.82    &    $-$7.75 \\
HD 168746  & 13660	& 2456226.63	& 0.590   &  9.96  & $<$3.17       & $<$0.32    & $<$1.26   & 42.73    & $<$26.59              & 7.95 & 33.60    & $<$$-$7.01 \\
HD 178911B & 13659	& 2455933.36	& 0.914   &  9.94  &    20.88      &    2.10    &    8.29   & 42.59    &    27.41$\pm$0.11     & 7.97 & 33.60    &    $-$6.19 \\
HD 187123  & 13658	& 2455970.76	& 0.465   &  9.76  & $<$3.15       & $<$0.32    & $<$1.26   & 48.26    & $<$26.70              & 7.83 & 33.75    & $<$$-$7.05 \\
HD 188015  & 13667	& 2456171.39	& 0.872   & 12.92  &    19.73      &    1.53    &    6.04   & 57.01    &    27.52$\pm$0.14     & 8.24 & 33.74    &    $-$6.22 \\
HD 204313  & 13669	& 2456222.07	& 0.786   &  9.66  & $<$3.12       & $<$0.32    & $<$1.26   & 47.37    & $<$26.69              & 7.99 & 33.67    & $<$$-$6.98 \\
51 Peg     & 10825\tablenotemark{a}	& 2454806.96	& 0.746   &  4.92  &    7.94       &    1.61    &    6.35   & 15.61    &    26.42$\pm$0.21     & 5.45 & 33.72    &    $-$7.30 \\[-15pt]
\enddata

\tablecomments{Table is ordered by RA. Column details: HJD is the
  heliocentric Julian Date at the beginning of the observation; $\phi$
  is the radial-velocity orbital phase relative to our line of sight;
  Exp is the effective exposure in ks; Cts is the net counts within a
  2$''$ aperture centered on the object coordinates at the time of the
  observation, or else the 95\% confidence upper limit; Rate is the
  net counts per ks; Flux is the unabsorbed 0.2--2~keV flux for a
  mekal model with $\log{T}=7$; Dist is the distance in parsecs
  calculated from {\it Hipparcos} parallax; $L_{\rm X}$ is the
  0.2--2~keV X-ray luminosity in erg~s$^{-1}$ (expressed as a
  logarithm).}  \tablenotetext{a}{From archival {\it Chandra\/} data;
  see $\S$2.2. All other {\it Chandra\/} observations are from our
  Cycle 13 program.}  \tablenotetext{b}{HD 134987 is formally a
  detection at $>$90\% confidence with our conservative methodology
  but is found by {\tt wavdetect}; see $\S$2.2.}
\end{deluxetable*}

The targeted stars are optically bright, with a median $v=7.8$. For
effective temperatures of 5500--6000~K, these magnitudes are expected
to result in approximately one photoelectron per pixel per standard
3.2~s frame exposure registering for the S3 chip near the
aimpoint. Each excess photoelectron shifts the bias level by 3.4~eV,
slightly altering the observed \hbox{X-ray} energy for a given
event. However, for these observations the total number of
\hbox{X-ray} counts is too low to conduct spectral analysis, so this
slight contamination has no impact on the derived fluxes (i.e.,
subarray binning was not required). The optical photoelectrons cannot
themselves register as an \hbox{X-ray} event, since the low energy
cutoff we use of 0.15~keV is $\sim$44 photoelectrons, corresponding to
stars 4 magnitudes brighter.

The 0.15--2~keV images of the targets are provided in
Figure~3. Because the targets are relatively nearby, it is necessary
to take proper motions into account. The J2000 coordinates were
adjusted to the epoch of the {\it Chandra\/} observations using the
proper motions measured by {\it Hipparcos\/} (van Leeuwen 2007). The
red and green circles show $r=2''$ apertures at the J2000 and updated
positions, respectively. Source detection is assessed through running
{\tt wavdetect\/} at a significance thresshold of $10^{-6}$ over the
S3 chip and separately by calculating whether the number of observed
counts is greater than could arise from background fluctuations at
95\% confidence (per the Bayesian-derived tables given in Kraft et
al.~1991). Five sources are clearly detected: HD 188015, HD 107148, HD
102117, HD 49674, and HD 178911B. Five sources are clearly not
detected: HD 204313, HD 28185, HD 6434, HD 168746, and HD 187123. One
source is borderline: HD 134987 has three counts and is formally a
detection at 90--95\% confidence using the criteria of Kraft et
al.~(1991), but those three counts are angularly concentrated within
$\sim$1.2$''$ (the on-axis PSF) and it is confirmed as a detection by
{\tt wavdetect}. One source is confused: HD 30177 has a clear
detection but that source is $\sim4''$ distant from the expected
target location; we conservatively consider this to be an unrelated
object, most plausibly a wide-orbit companion.

X-ray counts were extracted from the indicated 2$''$ apertures with
the (nearly neglible) background estimated from nearby source-free
regions. Net count rates were converted to unabsorbed fluxes using
PIMMS\footnote{{\tt http://cxc.harvard.edu/toolkit/pimms.jsp}} for a
fixed plasma/MEKAL model with three components at temperatures of
1~MK, 3~MK, and 10~MK ($\log{T}=6.0$, 6.5, and 7.0 or 0.09, 0.27, and
0.86 keV, respectively), with relative flux normalizations of
1:1:1. Solar abundances and an $N_{\rm H}=10^{18}$~cm$^{-2}$ were
assumed. This provides energy conversion factors between net rates and
fluxes similar to those adopted by previous studies of star-planet
interaction; for example, while we use 0.2--2~keV fluxes throughout,
the {\it ROSAT\/} PSPC and HRI rates would translate into 0.1--4.5~keV
fluxes through conversion by a factor of $5.6\times10^{-12}$ and
$2.9\times10^{-11}$, comparable to the factors of $6.5\times10^{-12}$
and $2.8\times10^{-11}$ adopted by Kashyap et al.~(2008). This model
also approximately reproduces the relative counts observed within the
0.2--0.45, 0.45--0.75, and 0.75--2.0~keV bands in {\it XMM-Newton\/}
observations of solar-type stars (Poppenhaeger et al.~2010). The
conversion factor from {\it Chandra\/}/ACIS-S (Cycle 13) net count
rates to unabsorbed flux (both over 0.2--2~keV) is then
$5.59\times10^{-12}$. Minor modifications in these parameters would
alter the resulting fluxes by less than the statistical
errors. \hbox{X-ray} luminosities are calculated for the Hipparcos
distances (van Leeuwen 2007), as for the bolometric
luminosities. X-ray luminosities are expressed as logarithms
throughout and given in units of erg~s$^{-1}$. We add an uncertainty
of 20\% to the statistical errors on the X-ray luminosities derived
from the {\it Chandra\/} observations to account for inaccuracies in
counts-to-flux conversion from using a fixed spectral model.

X-ray luminosities for the remainder of the full FGK MS sample are
recalculated from the net count rates given in Poppenhaeger et
al.~(2010)\footnote{Three stars from the sample in Poppenhaeger et
  al.~(2010) are not listed in the EOD; two are evolved and so would
  be excluded as non-MS anyway (HD 62509 and HD 27442) while one is no
  longer considered planet-bearing (HD 20367; Wittenmyer et
  al.~2009).} for stars observed by {\it XMM-Newton\/} (including
those from Kashyap et al.~2008), or else from {\it ROSAT\/}
coverage. For objects lacking {\it Chandra\/} or {\it XMM-Newton\/}
$L_{\rm X}$ values, we searched the RASS bright and faint source
catalogs as well as source catalogs generated from pointed PSPC and
HRI observations, prioritizing the latter. Net 0.1--2.5~keV {\it
  ROSAT\/} count rates were converted to 0.2--2~keV fluxes by factors
of 3.95$\times10^{-12}$ and 2.02$\times10^{-11}$ for the PSPC and HRI
detectors, respectively, while the energy conversion factor for {\it
  XMM-Newton\/} observations with the $pn$ detector and the
thick/medium/thin optical blocking filter is
3.90/2.66/2.43$\times10^{-12}$. The {\it XMM-Newton\/} conversion
factors were calculated using XSPEC and for the fixed spectral model
provide good agreement with the luminosities calculated by
Poppenhaeger et al.~(2010), with a median offset of 0.03 dex and a
scatter of 0.10 dex.\footnote{This is disregarding HD~195019 for which
  Poppenhaeger et al.~(2010) give a distance of 20~pc whereas we take
  38.5~pc from the parallax.} Approximate $L_{\rm X}$ upper limits
were estimated for undetected stars from the typical RASS faint source
catalog limit of $10^{-13 }$~erg~s$^{-1}$~cm$^{-2}$. In two cases we
use literature values for $L_{\rm X}$ that are based on the median of
multiple high-quality observations: $\upsilon$ And has {\it Chandra\/}
coverage described in Poppenhaeger et al.~(2011), while HD~179949 has
{\it XMM-Newton\/} coverage described in Scandariato et al.~(2013).

\section{Testing planetary correlations with X-ray luminosity}

\subsection{Coronal activity in solar analogs}

We tested for a statistically-significant correlation within the 23
solar analogs with X-ray coverage using the Kendall $\tau$ statistic
implemented within ASURV\footnote{{\tt
    http://astrostatistics.psu.edu/statcodes/asurv}} (Feigelson \&
Nelson~1985), which provides more accurate results than a Spearman
correlation test for small sample sizes. This is a non-parametric
ranking test that does not assume any particular functional
relationship between the variables. \hbox{X-ray} upper limits are
accounted for, but measurement errors are not; all points are equally
weighted. We considered four different potential interaction scalings:
$L_{\rm X}$ and $L_{\rm X}/L_{\rm bol}$ as activity indicators, versus
$M_{\rm P}/a^{2}$ and $1/a$ as interaction-strength proxies
(Figure~4). The values of $\tau$ are 0.28 and 0.21 for $L_{\rm X}$ and
$L_{\rm X}/L_{\rm bol}$ as a function of $M_{\rm P}/a^{2}$, and 0.42
and 0.35 as a function of $1/a$. These correspond to probabilities for
no correlation of 78\%, 83\%, 67\%, and 73\%; the null hypothesis
cannot be rejected, and by this metric the solar analogs show no
significant evidence for increased X-ray luminosity in hot Jupiter
systems.

We next tested for a statistically-significant correlation by
assessing whether a positive slope is present in a linear fit (using
logarithmic quantities). We use the Bayesian linear regression code of
Kelly (2007), which accounts for both measurement errors and upper
limits. The resulting best-fit values are reported as the median of
10000 draws from the posterior distribution, with uncertainties
corresponding to 1$\sigma$ given as the standard deviation. The
preferred slopes are near zero, specifically $\beta = -0.05\pm0.23,
-0.06\pm0.24$ and $\beta = 0.01\pm0.29, 0.02\pm0.30$ for $L_{\rm X}$,
$L_{\rm X}/L_{\rm bol}$ as a function of $M_{\rm P}/a^{2}$ and
$1/a$. Kashyap et al.~(2008) report an enhancement by a factor of
$\sim$4 in X-ray emission for close-in planets, of which they estimate
$\sim$2$^{+2}_{-0.7}$ is due to star-planet interaction, with the
remainder attributable to selection effects. The separation between
the putative weakly interacting and strongly interacting systems is
$\sim$1.7 dex in $M_{\rm P}/a^{2}$; a slope of 0.3 would then produce
an increase of 0.5 dex (or $\sim$3) in \hbox{X-ray} luminosity, so we
take the probability of $\beta>0.3$ as the likelihood that interaction
is present. The close-in and distant systems in $1/a$ are separated by
$\sim$1.0 dex, so here we consider $\beta>0.5$ to provide positive
evidence of interaction. The posterior distribution of $\beta$ is
shown in Figure~5; in all cases, slopes above these cutoff values are
excluded at $\simgt94$\% confidence.\footnote{Confidence estimates
  here and throughout are given as the percentage of draws from the
  posterior distribution satisfying the relevant condition, which is
  more accurate than extrapolating from a preferred parameter value
  and its estimated 1$\sigma$ errors.} The insets show 500 draws from
the posterior distribution for the intercepts and slopes. The linear
regression slopes likewise and independently do not produce
significant evidence for star-planet interaction in the solar analogs.

\begin{figure}
\includegraphics[scale=0.44]{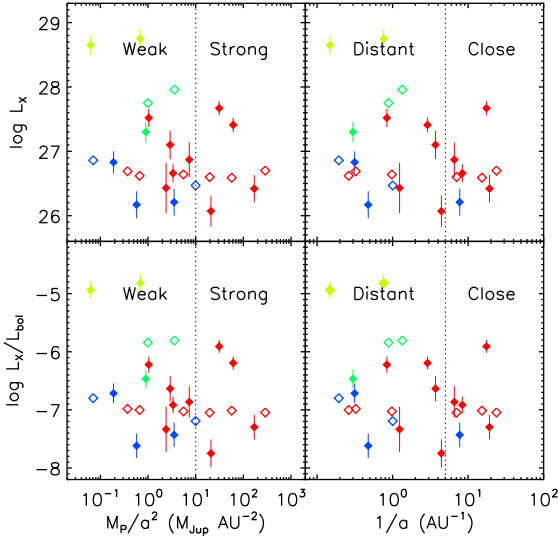} \figcaption{\small
  Distribution of coronal activity ($\log{L_{\rm X}}$, top, and
  $\log{L_{\rm X}/L_{\rm bol}}$, bottom) versus two proxies for
  interaction strength ($\log{M_{\rm P}/a^{2}}$, left, and
  $\log{1/a}$, right) for the subsample of solar analogs. Red, blue,
  and green colors mark {\it Chandra\/}, {\it XMM-Newton\/}, and {\it
    ROSAT\/} measurements (the light-green points are excluded as
  atypically active). Filled symbols are \hbox{X-ray} detections and
  open symbols are upper limits. Vertical dashed lines delineate
  weakly versus strongly interacting systems at $M_{\rm P}/a^{2}=10
  M_{\rm Jup}$~AU$^{-2}$ and close versus distant planets at
  $a=0.2$~AU.}
\end{figure}

We also tested the significance of our result by conducting
simulations, to assess whether 23 objects is sufficient to identify or
exclude interaction. For each trial, we fixed the values of $M_{\rm
  P}/a^{2}$ at those for the observed solar analogs, and then
generated $L_{\rm X}/L_{\rm bol}$ for two hypothetical distributions:
first, a normal distribution about $-7$ with scatter 0.7 dex (matching
the preferred value fit to the data) and no dependence upon $M_{\rm
  P}/a^{2}$, and second, a normal distribution about the line $L_{\rm
  X}/L_{\rm bol}$ = $-7 + 0.3\times(\log{M_{\rm P}/a^{2}}-0.6)$, again
with 0.7 dex scatter. For the second model the slope is selected such
that the increase in $L_{\rm X}/L_{\rm bol}$ is 0.5 dex over the 1.7
orders of magnitude separating hypothetically weakly and strongly
interacting systems, which matches the $\sim$2 (4) factor of increased
\hbox{X-ray} activity found by Kashyap et al.~(2008) after (prior) to
controlling for selection effects. In only 7\% of cases with no input
interaction is the best-fit slope fit to the random realizations
greater than 0.3 (i.e., $\sim$7\% Type I error). In only 4\% of cases
with a signficant input interaction is the best-fit slope less than
zero (i.e., $\sim$4\% Type II error). These simulations indicate that
our sample of solar analogs is sufficiently large in size and dynamic
range to guard against randomly pathological distributions and
verifies that the results of the correlation and linear regression
tests are secure.

\begin{figure}
\includegraphics[scale=0.48]{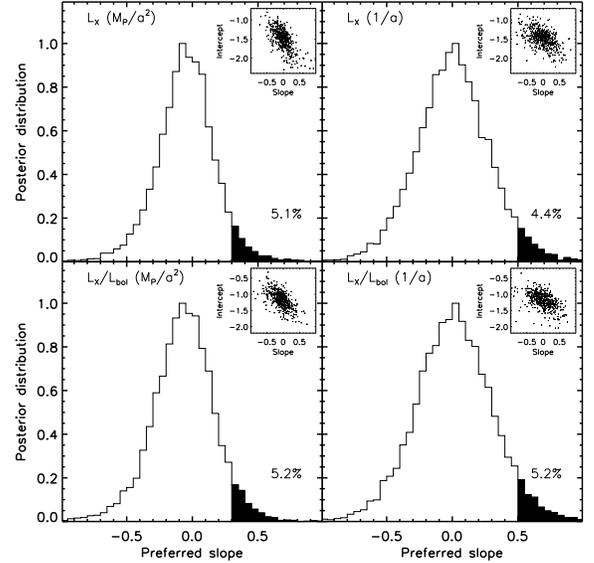} \figcaption{\small
  Histograms of the posterior distribution of slopes for linear fits
  (to logarithmic values) to the solar analogs subsample for the
  quantities shown in Figure~4 (panels identically ordered). In all
  cases the median slope is consistent with zero (i.e., no
  interaction). The filled segments mark $\beta>0.3$ (left) and
  $\beta>0.5$ (right) which would produce an enhancement by $\sim$3 in
  \hbox{X-ray} luminosity across the 1.7 and 1.0 orders of magnitude
  separating typical weakly versus strongly (left) and close versus
  distant (right) systems; such an enhancement is ruled out at
  $\simgt94$\% confidence. The insets show 500 points drawn from the
  intercept and slope posteriors.}
\end{figure}

\begin{figure}
\includegraphics[scale=0.49]{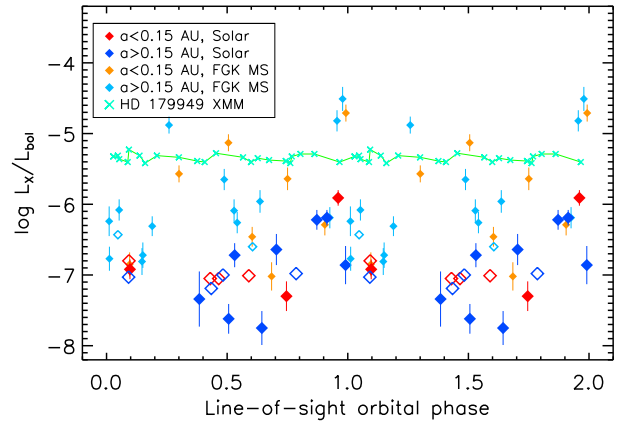} \figcaption{\small The
  line-of-sight orbital phase versus $L_{\rm X}/L_{\rm bol}$ for solar
  analogs hosting close-in ($a<0.15$~AU, red) and distant
  ($a>0.15$~AU, blue) planets. For context, non-RASS single-pointing
  \hbox{X-ray} measurements from FGK MS stars are also shown (orange
  and cyan for close-in and distant planets). X-ray limits are
  indicated with open symbols. The {\it XMM-Newton\/} monitoring
  campaign on HD~179949 from Scandariato et al.~(2013) is also
  shown. Phases are calculated from the orbital parameters in the EOD
  (derived from fitting RV measurements) and the HJD at the beginning
  of the observation, and zero corresponds to the planet crossing on
  the near side of the star (although note these are not transiting
  systems). The data are repeated with the phase offset by one.}
\end{figure}

Finally, we consider the orbital phase at which the {\it Chandra\/}
observations of solar analogs were conducted. It is possible that
magnetic star-planet interaction could occur preferentially near a
particular phase. For example, a reconnection-induced hotspot slightly
leading the sub-planetary point on the stellar surface would cross the
line of sight prior to the planet, at a phase slightly less than
unity. Based on \ion{Ca}{2} H and K variability in HD 179949 and
$\upsilon$ And, Shkolnik et al.~(2008) identified $\phi\sim0.8$ as an
orbital phase at which star-planet interaction is preferentially
manifested. On the other hand, Pillitteri et al.(2010, 2011, 2014a)
find \hbox{X-ray} activity in HD~189733 near $\phi\sim0.5$ (i.e., when
the planet is behind the star). A planet-induced coronal hotspot being
dragged across the stellar surface should be visible over $\sim$0.5 of
the planetary orbit, with the projected surface area peaking along the
line of sight. We determined the heliocentric Julian date at the
midpoint of the {\it Chandra\/} observation, and converted this to the
line-of-sight orbital phase using the orbital parameters in the EOD
(derived from fitting RV measurements). Figure~6 shows $\phi$ versus
$L_{\rm X}/L_{\rm bol}$ for solar analogs hosting close-in (red) and
distant (blue) systems. For context, we also plot \hbox{X-ray}
luminosities from single-pointing non-RASS coverage of FGK MS stars,
generally from {\it XMM-Newton\/} observations. We also show the {\it
  XMM-Newton\/} monitoring campaign on HD 179949 from Scandariato et
al.~(2013; their Table~6); for extensive \hbox{X-ray} coverage and
discussion of $\upsilon$ And, HD~179949, or HD~189733 we refer the
reader to Poppenhaeger et al.~(2011), Saar et al.~(2008) plus
Scandariato et al.~(2013), and Poppenhaeger et al.~(2013) plus
Pillitteri et al.~(2010, 2011, 2014a), respectively. Taking both
detection and limits into account, there are no particular phases at
which the close-in systems show systematic \hbox{X-ray} enhancements
(or large positive-only scatter). While the number of points is not
large enough to draw definitive conclusions, there is no evidence from
these observations of a preferred phase at which planet-induced
enhancements routinely occur in hot Jupiter systems.

\begin{figure*}
\includegraphics[scale=0.96]{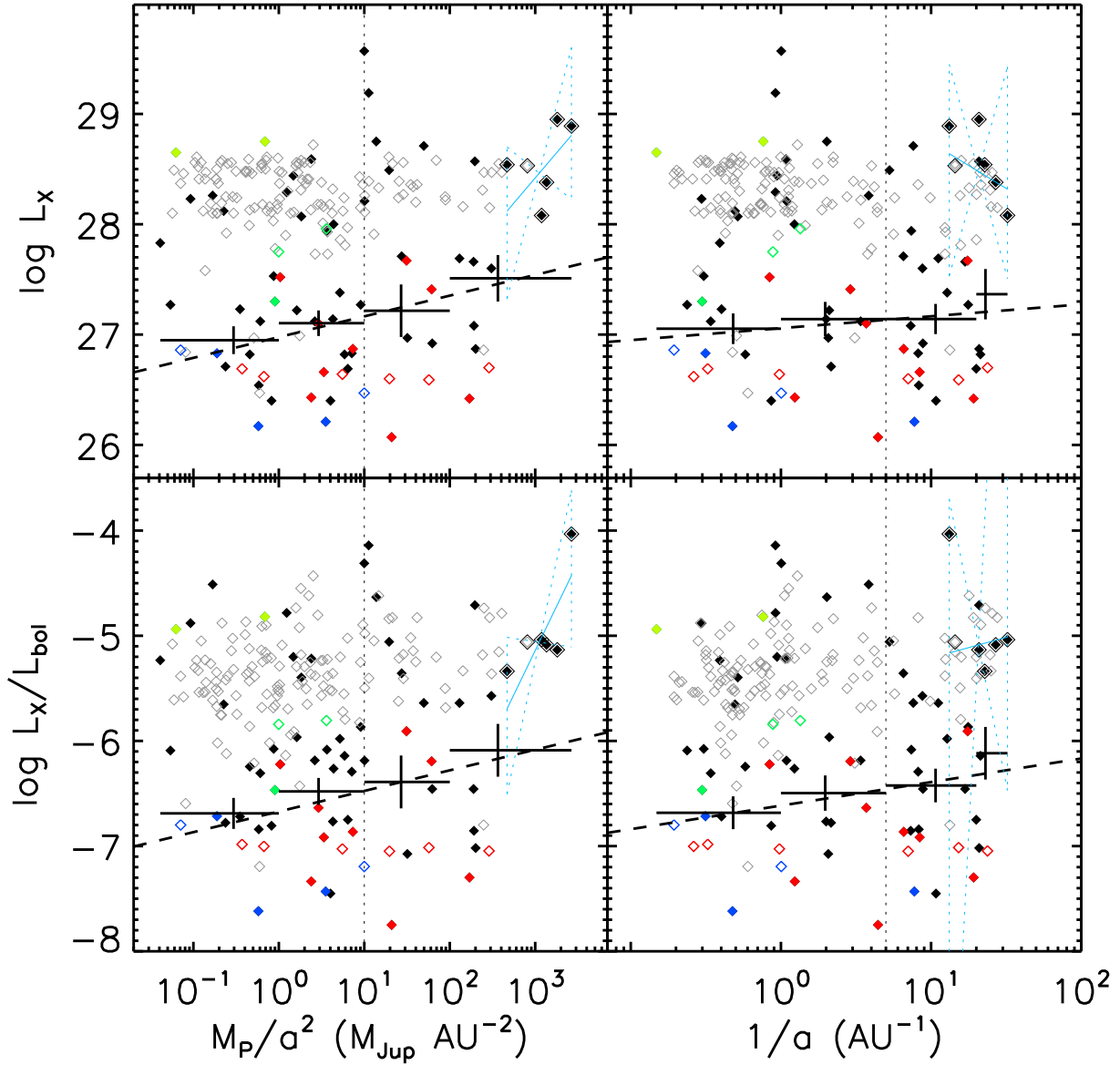} \figcaption{\small
  Distribution of coronal activity ($\log{L_{\rm X}}$, top, and
  $\log{L_{\rm X}/L_{\rm bol}}$, bottom) versus two proxies for
  interaction strength ($\log{M_{\rm P}/a^{2}}$, left, and
  $\log{1/a}$, right) for the full sample of main sequence
  stars. Colored symbols are as in Figure~4, and the black filled and
  gray open diamonds are X-ray detections and upper limits,
  respectively. The double-diamond symbols mark ``extreme'' systems
  (see $\S$3.2 and $\S$4.3) for which separate fits and corresponding
  errors on the slope are plotted as solid and dotted cyan lines. The
  black crosses show Kaplan Meier mean values (taking upper limits
  into account) within the bins indicated by the horizontal arms, with
  uncertainties as shown by the vertical arms. The dashed line is a
  linear regression to the full dataset taking both upper limits and
  measurement uncertainties into account. Vertical dotted lines
  separate strongly from weakly interacting systems as in Figure~4.}
\end{figure*}

\subsection{Coronal activity in main sequence FGK stars}

For the full main sequence sample, there are sufficient stars to
compare $L_{\rm X}$ and $L_{\rm X}/L_{\rm bol}$ within four discrete
bins of $M_{\rm P}/a^2$ and $1/a$. The Kaplan Meier mean is computed
within ASURV, taking X-ray upper limits into account, and the X-ray
luminosities are centered to $L_{\rm X}-28$ and $L_{\rm X}/L_{\rm
  bol}+6$. The results are plotted as large crosses in Figure~7, with
the horizontal bars indicating bin boundaries and vertical bars
indicating the error on the mean within that bin. While the
uncertainties are large, $L_{\rm X}$ increases by $0.56\pm0.25$ dex
from the lowest to the highest $M_{\rm P}/a^2$ bin (with means, as
logarithms, of $-0.53$ and 2.56 $M_{\rm Jup}$~AU$^{-2}$, respectively,
a separation of $\sim$3 dex), and $L_{\rm X}/L_{\rm bol}$ similarly
increases by $0.60\pm0.29$ across this same interval. The difference
is less pronounced for $1/a$ as a function of \hbox{X-ray} luminosity;
from the most distant to the most close-in bins (with means, as
logarithms, of $-0.32$ to 1.36 AU$^{-1}$, respectively, a separation
of $\sim$1.7 dex), the increase is $0.31\pm0.27$ and $0.57\pm0.29$ for
$L_{\rm X}$ and $L_{\rm X}/L_{\rm bol}$, respectively.

Testing the main-sequence sample for a correlation with ASURV, the
Kendall $\tau$ values are 2.37 and 1.91 for $L_{\rm X}$ and $L_{\rm
  X}/L_{\rm bol}$ as a function of $M_{\rm P}/a^2$, which correspond
to probabilities for no correlation of 1.8\% and 5.6\%; the null
hypothesis is rejected at $\simgt2\sigma$, confirming the increase in
X-ray luminosity for hot Jupiters in the full sample, in contrast to
the results for the solar analogs alone. On the other hand, the
Kendall $\tau$ values are 1.09 and 1.55 for $L_{\rm X}$ and $L_{\rm
  X}/L_{\rm bol}$ as a function of $1/a$, which correspond to
probabilities for no correlation of 28\% and 12\%; here the null
hypothesis cannot be rejected.

Qualitatively similar results are obtained for fitting $L_{\rm X}$ or
$L_{\rm X}/L_{\rm bol}$ as a function of $M_{\rm P}/a^2$ or $1/a$
(again with all quantities expressed as logarithms and centered). The
best-fit linear relations (calculated with the Bayesian IDL routine of
Kelly 2007) are overplotted in Figure~7 as dashed black lines. This
methodology is completely independent of the mean values found within
bins with ASURV, but it may be observed that there is good agreement
in the trends. Specifically, the preferred slopes, with $1\sigma$
errors, are $0.19\pm0.08$ and $0.20\pm0.08$ for $L_{\rm X}$ and
$L_{\rm X}/L_{\rm bol}$ as a function of $M_{\rm P}/a^2$, and
$0.11\pm0.13$ and $0.22\pm0.15$ as a function of $1/a$, with
corresponding likelihood over the full posterior for the slope to be
greater than zero of 99.6\%, 99.3\%, 79.8\%, and 93.7\%. (These
preferred slopes are still significantly less than the values of 0.3
and 0.5 that would correspond to a typical increase in \hbox{X-ray}
emission by a factor of $\sim$3 for $M_{\rm P}/a^2$ and $1/a$,
respectively.)

The significance by which the preferred slope exceeds zero is strongly
dependent upon a handful of extreme systems. Specifically, if the six
systems (HIP 14810, HD 73256, tau Boo, HD 162020, HD 179949, and HD
189733) with $M_{\rm P}/a^2>450$~$M_{\rm Jup}$~AU$^{-2}$ were to be
omitted, preferred slopes for all four fits would be consistent with
zero ($0.06\pm0.09$, $0.04\pm0.09$, $-0.10\pm0.14$,
$0.03\pm0.15$). HIP 14810 has only a loose X-ray limit, but the other
five extreme systems are X-ray luminous and also chromospherically
active, with $R_{\rm HK}^{'}\simgt-4.7$. For completeness, we also fit
linear relations to the extreme systems only (solid cyan lines in
Figure~7), but because there are only six points the uncertainties in
the slopes (dotted cyan lines) are quite large and permit slopes of
zero, preventing us from drawing any definitive conclusions.

An older main-sequence star will have lower activity than an otherwise
similar but younger counterpart, independent of planetary
properties. Because the inferred signatures of star-planet interaction
upon the chromospheric emission cores in the \ion{Ca}{2} H and K lines
is at the few percent level (Shkolnik et al.~2008), the value of
$R^{'}_{\rm HK}$ for a given star should be nearly or entirely
independent of planetary properties (e.g., Canto Martins et al.~2012;
see further discussion in Section 5). We search for excess X-ray
luminosity beyond that related to the intrinsic stellar activity by
removing the dependence of $L_{\rm X}/L_{\rm bol}$ upon $R^{'}_{\rm
  HK}$, as parameterized for main-sequence stars by Mamajek \&
Hillenbrand (2008). The resulting preferred slopes are consistent with
zero at the $\simlt1\sigma$ level ($0.03\pm0.06$, $0.03\pm0.06$,
$0.09\pm0.12$, and $0.09\pm0.11$). The preferred intercepts are
modestly negative ($-0.23$, $-0.24$, $-0.27$, $-0.26$, all $\pm$0.09),
perhaps reflective of the (selected) lower activity of radial-velocity
targeted stars compared to those in the field (see also discussion in
Shkolnik 2013). If only the most chromospherically active stars are
considered, there are 31 stars with $R^{'}_{\rm HK}>-4.8$, of which 16
have X-ray detections, and the preferred slopes for this subset
against $L_{\rm X}$ or $L_{\rm X}/L_{\rm bol}$ are consistent with
zero at the 1$\sigma$ level.

An alternative method of testing whether a potential correlation
between two variables could be related to a third variable is given by
the Kendall partial tau test. This has been implemented for data
including censoring by Akritas and Seibert (1996) and we use their
methodology to test whether the correlation between $L_{\rm X}$ and
$M_{\rm P}/a^{2}$ remains significant when controlling for $R^{'}_{\rm
  HK}$. Kelly et al.~(2007) identify instances in which this test can
produce misleading results, but our sample is not subject to the
strong multiple variable correlations that can be potentially
problematic, and we use this test only as an additional check. For
reference, with the third variable also equal to $M_{\rm P}/a^{2}$ but
summed with a random Gaussian of standard deviation 0.001,
$\tau_{1,3}=0.9998$ and $\tau_{12,3}=5.6\times10^{-3}$, with
$\sigma=7.5\times10^{-3}$, properly identifying the third variable as
relevant; in contrast, controlling for a uniformly zero third variable
produces $\tau_{1,3}=0.015$ and $\tau_{12,3}=0.053$, with
$\sigma=0.026$, properly rejecting the third variable as
relevant. Controlling for $R^{'}_{\rm HK}$, $\tau_{1,2}=0.038$,
$\tau_{1,3}=0.058$, $\tau_{2,3}=0.12$, and $\tau_{12,3}=0.031$, with
$\sigma=0.023$. This indicates that the null hypothesis cannot be
rejected and so the influence of $R^{'}_{\rm HK}$ upon the $L_{\rm X}$
versus $M_{\rm P}/a^2$ correlation cannot be ruled out, consistent
with the regression results. For completeness, we verified that a
partial correlation with either distance or stellar temperature is
rejected. These results are consistent with and reinforce those
obtained from the two-variable correlation and linear regression
tests.

While magnetic star-planet interaction might provide one explanation
for these results, at least for the most extreme systems, it is
necessary to consider also selection biases as well as planet and
stellar evolution effects.

\section{Discussion of X-ray results}

The lack of a systematic increase in coronal activity in hot Jupiter
systems in our controlled investigation of solar analogs indicates
that any magnetic star-planet interaction must be either uncommon or
of low efficiency in these systems. These results limit the immediate
utility of star-planet interaction as a general probe of exoplanet
magnetic field strengths (the complementary extreme-system study of
WASP-18 by Miller et al.~2012 reached similar conclusions). We next
discuss various factors that may influence comparative X-ray emission
across the full FGK MS sample, and specifically consider radial
velocity sensitivity bias, binarity, star-planet interaction in
extreme systems, and cumulative tidal effects.

\subsection{Radial velocity sensitivity bias}

The full sample of main-sequence FGK planet-hosting stars is
susceptible to selection biases; in particular, weakly interacting
systems may only be detectable around inactive stars, resulting in a
deficit of systems with low-mass, distant planets and high $L_{\rm X}$
values and inducing a spurious correlation that must be removed prior
to evaluating potential star-planet interaction signatures. This
incompleteness is apparent in a diagram of the velocity semi-amplitude
versus the residual "jitter" noise (Figure~8). Here $K$ is primarily
influenced by the planetary properties (the range of stellar masses is
less than that in period and planetary mass) while $RMS$ reflects the
intrinsic activity of the parent star. There is a dearth of points at
low $K$ and high $RMS$ values due to sensitivity limitations in radial
velocity searches.

This bias results in relatively fewer systems at low $M_{\rm P}/a^{2}$
and high $L_{\rm X}$ values. We confirm that a significant selection
bias is present in our data through testing the scaling of the
velocity semi-amplitude $K$ with distance, and find that a positive
correlation is present at $>3\sigma$ whereas in a complete sample no
dependence would be expected. (Excluding the six most extreme systems
with $M_{\rm P}/a^2>450$~$M_{\rm Jup}$~AU$^{-2}$ does not remove the
correlation with $K$).

Restricting consideration to a portion of the $K$-$RMS$ plane with
complete coverage, defined here as $20<K\le500$~m~s$^{-1}$ and
$2<RMS\le15$~m~s$^{-1}$ (the solid square in Figure~8), produces a
subsample of 110 systems, of which 30 have \hbox{X-ray} detections,
for which no correlation (slope $0.017\pm0.028$) is now present
between $K$ and distance. Four of the five \hbox{X-ray} detected
extreme systems with $M_{\rm P}/a^{2}>450 M_{\rm Jup}$~AU$^{-2}$ are
included in this subset; Tau Boo is excluded with
$RMS>15$~m~s$^{-1}$. This subsample does show a significant dependence
of $L_{\rm X}$ or $L_{\rm X}/L_{\rm bol}$ upon $M_{\rm P}/a^{2}$
($\simgt99$\% probability that the slope $>0$) and the preferred
slopes are similar to those found for the full sample (albeit less
tightly constrained), which suggests that the RV-sensitivity selection
bias is not the sole driver associating enhanced \hbox{X-ray} emission
with hot Jupiter systems. If instead the subsample is restricted to
$20<K\le200$~m~s$^{-1}$ and $2<RMS\le15$~m~s$^{-1}$ (the dotted line
in Figure~8; 95 systems, of which 24 have \hbox{X-ray} detections),
four of the five extreme systems are now excluded as having
$K>200$~m~s$^{-1}$ (HD~179949 is retained). Here again there is no
correlation between $K$ and distance, but there is still suggestive
evidence for a correlation between $L_{\rm X}$ or $L_{\rm X}/L_{\rm
  bol}$ and $M_{\rm P}/a^{2}$ ($\simgt91$\% probability that the slope
$>0$, dropping to 80\% if HD~179949 is excluded). Removing the
dependence of $K$ upon distance through selection of a $K$-$RMS$
complete subsample does not eliminate the trend of increasing $L_{\rm
  X}$ or $L_{\rm X}/L_{\rm bol}$ toward greater $M_{\rm P}/a^{2}$,
unless the most extreme systems are all deliberately excluded from the
subsample.

We emphasize that the identification of a selection bias within a
sample does not necessarily indicate that any correlation with
planetary properties is due to that bias. In particular, since $K$
scales with $M_{\rm P}/a^{0.5}$, most proxies for interaction
strength, including the $M_{\rm P}/a^{2}$ preferred here, will
correlate with $K$.

\begin{figure}
\includegraphics[scale=0.47]{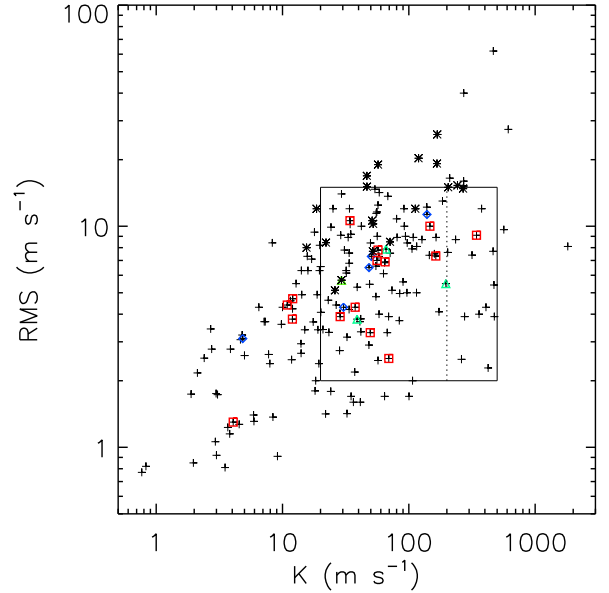} \figcaption{\small Velocity
  semi-amplitude $K$ versus RMS velocities about a best-fit Keplerian
  for the sample of FGK main-sequence planet-hosting stars. The
  paucity of points at large RMS and small $K$ values primarily
  results from the sensitivity incompleteness of radial velocity
  detections. The square defines an RV-complete subsample.}
\end{figure}

\subsection{Binary systems}

Planet-hosting stars are sometimes present in binary systems
($\sim$10--20\%; Raghavan et al.~2006; Roell et al.~2012). Binarity
increases intrinsic \hbox{X-ray} activity even in long-period systems,
perhaps related to initial formation (Pye et al.~1994). In addition,
any unresolved \hbox{X-ray} emitting secondary companions would
inflate the apparent \hbox{X-ray} luminosity of the planet-hosting
primary; this is more of a concern for the {\it XMM-Newton\/} and
particularly {\it ROSAT\/} observations, which lack the angular
resolution of {\it Chandra\/}, but note that 25\% contamination would
only increase $L_{\rm X}$ by 0.1 dex. Among our new {\it Chandra\/}
targets, HD 188015 and HD 178911B are in wide binary\footnote{HD
  178911 is actually a triple, with AC having a separation of 0.1$''$
  (Raghavan et al.~2006).} star systems (Raghavan et al.~2006) with
projected component separations of 16$''$ and 13$''$ (corresponding to
790 and 680 AU), respectively. Both companions are detected in the
{\it Chandra\/} observations (Figure~3). The dynamics and evolution of
planetary systems are sensitive to binarity; for example, Kaib et
al.~(2013) demonstrate that outer exoplanets within wide binaries can
be destabilized as the system responds to Milky Way tidal forces and
passing stars, an effect that could artificially link (surviving) hot
Jupiter systems with binary (more active) stars.

We consider whether binarity corresponds with increased \hbox{X-ray}
activity in stars flagged as binary in the EOD (since these stars
possess sensitive RV measurements required to discover planets, this
assessment should also be complete with respect to close stellar
companions). The full sample of 198 planet-hosting main-sequence stars
contains 36 known binary systems. The Kaplan Meier mean values of
$L_{\rm X}-28$ are $-0.94\pm0.09$ and $-0.58\pm0.18$ for single stars
(46/162 detected) and binary (16/36 detected) systems, respectively,
calculated within ASURV. The mean values of $L_{\rm X}/L_{\rm bol}+6$
are $-0.56\pm0.10$ and $-0.30\pm0.17$. While the distributions of
$L_{\rm X}$ are marginally distinct (probability of 4.2\% using the
Peto \& Prentice test), the distributions of $L_{\rm X}/L_{\rm bol}$
are not inconsistent (Peto \& Prentice probability 25\%). The binary
systems also have, on average, modestly more extreme planets, with
mean $\log{M_{\rm P}/a^{2}}=0.92$ and mean $\log{1/a}=0.37$ versus
0.35 and 0.21 for the single stars. Among the most extreme hot Jupiter
systems, $\tau$ Boo and $\upsilon$ And are known binaries (Butler et
al.~1997), as is HD 189733 (Roell et al.~2012).

\subsection{Magnetic interaction in extreme systems}

In $\S$3.2 we found that a handful of extreme systems with $M_{\rm
  P}/a^{2}>450 M_{\rm Jup}/AU^{2}$ drive the correlation with $L_{\rm
  X}$ or $L_{\rm X}/L_{\rm bol}$ in the FGK MS sample. Extreme systems
are easier to detect in RV searches (see $\S$4.1), so a tendency
toward higher \hbox{X-ray} luminosities within this bin is not a
selection bias.\footnote{It should be noted that there are exceptions
  to this trend; most notably, WASP-18 is strongly tidally interacting
  but X-ray weak with a low $R_{\rm HK}^{'}$ (Miller et al.~2012;
  Pillitteri et al.~2014b), and the solar-type star XO-5 has
  serendipitous {\it Chandra\/} coverage that suggests it is at most
  moderately X-ray bright, with $L_{\rm X}/L_{\rm bol}<-5.3$ (these
  transit-detected systems are not in our sample).} Compared to the
solar analogs for which no evidence of magnetic star-planet
interaction is found, these extreme systems have similar masses but
smaller semi-major axes. In addition, their shorter orbital periods
produce faster rotation rates (for tidal locking at $a<0.15$~AU;
Bodenheimer et al.~2001), which would also increase the interaction
energy. The stellar magnetic fields are also likely to be higher,
based on the $R_{\rm HK}^{'}$ values; for example, HD~189733 has a
directly measured value of about 30~G (Fares et al.~2010), although
for HD 179949 it is only a few Gauss (Fares et al.~2012), comparable
to the Sun. We briefly highlight the properties of the
\hbox{X-ray}-detected extreme systems and summarize previous
phase-resolved studies that searched for or identified apparent
star-planet interaction.

HD 73256 is a solar-temperature ($T_{\rm eff}=5600$~K) but active
star, with $R_{\rm HK}^{'}=-4.5$ and $L_{\rm X}/L_{\rm
  bol}=-4.9$. Shkolnik et al.~(2005) find variation of the \ion{Ca}{2}
H and K cores to be modulated with the stellar rotation period (with a
flare observed near orbital phase 0.03), but note that the amplitude
of the stellar variability could dilute a planet-induced
signature. Shkolnik et al.~(2008) retain it as a candidate for
observable interaction.

Tau Boo is a hot ($T_{\rm eff}=6400$~K) and nearby ($d=15.6$ pc) star
that is moderately active, with $R_{\rm HK}^{'}=-4.7$ and $L_{\rm
  X}/L_{\rm bol}=-5.1$. (A more recent {\it Chandra\/} observation
resolves the secondary and finds a somewhat lower \hbox{X-ray} value
for the primary; Poppenhaeger \& Wolk 2013). Because the stellar
rotation period is identical or nearly so to the planetary orbital
period, identification of persistent interaction requires long-epoch
studies. Walker et al.~(2007) do indeed find that starspot activity is
concentrated near a fixed orbital phase. However, as noted by Shkolnik
et al.~(2008), tidal locking of the star to the planet would produce
low relative velocities between field lines, limiting the energy
available in magnetic interactions.

HD 162020 is a cool ($T_{\rm eff}=4800$~K) star that is categorized as
pre-main sequence in SIMBAD. Poppenhaeger \& Schmitt (2011)
consequently exclude it when examining the correlation found by Scharf
(2010). It is extremely X-ray luminous, with $L_{\rm
  X}=29.1$~erg~s$^{-1}$ and $L_{\rm X}/L_{\rm bol}=-3.8$, and also has
a very massive hot Jupiter, with $M\sin{i}=15.2 M_{\rm Jup}$. There is
no measurement of $R_{\rm HK}^{'}$ available in the literature.

HD 179949 is a hot ($T_{\rm eff}=6200$~K) star that is moderately
active, with $R_{\rm HK}^{'}=-4.6$ and $L_{\rm X}/L_{\rm
  bol}=-5.3$. It is the first identified candidate to display
star-planet interaction (Shkolnik et al.~2003) and has been
extensively investigated since, with possible \ion{Ca}{2} H and K
variability phased with the planet at some epochs but not others
(Shkolnik et al.~2008), and potential \hbox{X-ray} star-planet
synchronicity identified by Saar et al.~(2008). It is now clear that
the magnetic field configuration is complex (Fares et al.~2012) and so
a simple hotspot model might not be applicable. A recent large {\it
  XMM-Newton\/} plus optical study of HD~179949 did not find
significant evidence of star-planet interaction in either X-rays or
\ion{Ca}{2} H and K emission (Scandariato et al.~2013).

HD 189733 is a cool ($T_{\rm eff}=5000$~K) star that hosts a
transiting hot Jupiter. As such, it is among the most observed
exoplanet systems to date, with numerous deep campaigns at optical
(ground-based and {\it HST\/} and X-ray ({\it Chandra\/} and {\it
  XMM-Newton\/}) wavelengths. The star is active, with $R_{\rm
  HK}^{'}=-4.5$ and $L_{\rm X}/L_{\rm bol}=-4.8$. Pillitteri et
al.~(2010, 2011, 2014a) find cases of X-ray flaring near orbital phase
$\phi\sim0.5$ (i.e., when the planet is behind the star). Shkolnik et
al.~(2008) find \ion{Ca}{2} H and K emission variability to phase with
the star but suggest a residual signature of interaction near
$\phi\sim0.8$; on the other hand, Fares et al.~(2010) find no evidence
of magnetospheric interactions. Poppenhaeger et al.~(2013; also
Pillitteri et al.~2014a) find that the companion is not \hbox{X-ray}
bright and so the system age is likely older than would be inferred
from the activity of the primary.

\begin{figure}
\includegraphics[scale=0.47]{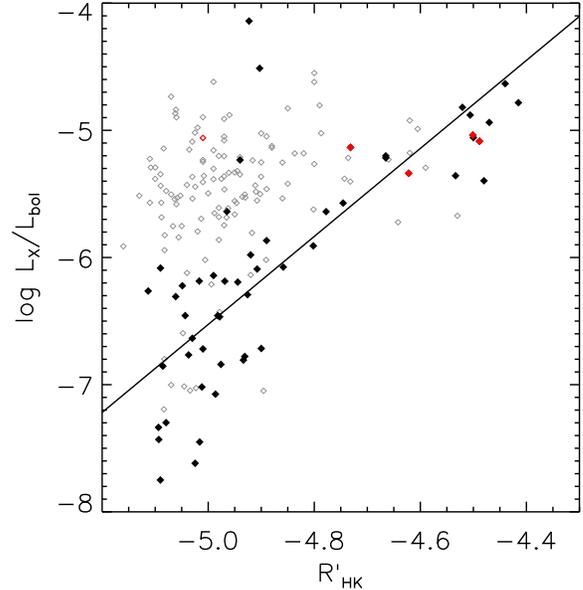} \figcaption{\small Coronal
  versus chromospheric activity for the sample of FGK MS stars with
  measured $R_{\rm HK}^{'}$ values. X-ray upper limits are plotted as
  open symbols, and extreme systems with $M_{\rm P}/a^2>450$~$M_{\rm
    Jup}$~AU$^{-2}$ are colored in red. The solid line shows the
  best-fit relation for $L_{\rm X}/L_{\rm bol}$ as a function of
  $R_{\rm HK}^{'}$ from Mamajek \& Hillenbrand (2008).}
\end{figure}

Based on phase-resolved studies of the most promising candidates, in
particular HD~179949, the evidence for star-planet interaction is
mixed; at best it seems the phenomenon is observable at select
epochs. We reiterate that these extreme systems are not X-ray luminous
relative to their chromospheric activity (Figure~9), which has been
suggested to be the case for star-planet interaction (e.g., Kashyap et
al.~2008). It is therefore possible that the most extreme systems are
more likely to be active, but for reasons unrelated to direct magnetic
interaction.

\subsection{Cumulative tidal effects in extreme systems}

Extreme systems are susceptible to cumulative tidal evolution. A
close-in gas giant can potentially spin-up its host star, temporarily
halting or reversing the usual decline in rotation and dynamo activity
with MS age, while gradually shrinking its orbit until infall (Jackson
et al.~2009; Debes \& Jackson 2010; see also Lanza \& Shkolnik 2014
for multi-planet possibilities). There is some observational support
for activity rejuvenation in binary systems; for CoRoT-2 and HD
189733, the hot Jupiter hosting primary is active while the companion
is \hbox{X-ray} quiescent (Schr{\"o}ter et al.~2011; Pillitteri et
al.~2011; Poppenhaeger \& Wolk 2013, 2014). In these cases $M/a^{3}$
is much larger for the planet than for the companion star, and the
lack of simultaneous activity rejuvenation in the companion implies
the tidal influence of the planet is more relevant. In addition,
$\tau$ Boo is apparently tidally locked to the orbit of its hot
Jupiter (Shkolnik et al.~2008) at a rapid rotation period of only
$\sim$3.2~d, strong evidence for planetary spin-up. At the same time,
extreme systems are tidally unstable on timescales that depend
sensitively on the orbital semi-major axis, eccentricity, planetary
composition, and stellar size. For example, WASP-18 has a 10 $M_{\rm
  Jup}$ planet in a 0.94 day orbit and an estimated age of $\sim$700
Myr (Hellier et al.~2009; Pillitteri et al.~2014b); the remaining
lifetime of WASP-18b is likely quite short, only $\sim$50 Myr (Hellier
et al.~2009).\footnote{WASP-18 experiences significant tidal effects
  from the planet (Arras et al.~2012) and may also be spun-up,
  although it is still inactive (see Miller et al.~2012, Pillitteri et
  al.~2014b, and references therein).} The relative paucity of hot
Jupiter systems with stellar ages above a few Gyr and semi-major axes
less than 0.05 AU is cited by Jackson et al.~(2009) as reflecting
tidal destruction on Gyr timescales. Subsequent age-rotation-activity
evolution (unaffected by any remaining outer planets) would then
decrease \hbox{X-ray} luminosity by $1-2$ orders of magnitude over the
MS lifetime (Ribas et al.~2005). If in fact extreme systems do not
survive beyond a few Gyr, then they are primarily present around
younger stars that are both inherently more active and susceptible to
cumulative tidal spin-up activity rejuvenation.

This interpretation is completely consistent with both the lack of
star-planet interaction in the solar analogs (which do not feature the
most extreme systems, and are screened to exclude active and likely
young stars) and the trend toward greater X-ray luminosities with
$M_{\rm P}/a^{2}$ in the full FGK MS sample. It may be that selection
biases (acting primarily at low $M_{\rm P}/a^{2}$) and cumulative
tidal influences (relevant primarily at high $M_{\rm P}/a^{2}$)
combine to mimic the statistical signature of star-planet
interaction. Of course, these effects do not rule out magnetic
recombination events producing enhanced activity,\footnote{It has also
  been suggested that magnetic reconnection events could act to hinder
  spin down (Lanza 2010) or could decay orbits (Strugarek et
  al.~2014).} but they do suggest that this type of interaction is not
required to explain the statistical trends in \hbox{X-ray} studies.

\begin{figure*}
\includegraphics[scale=0.85]{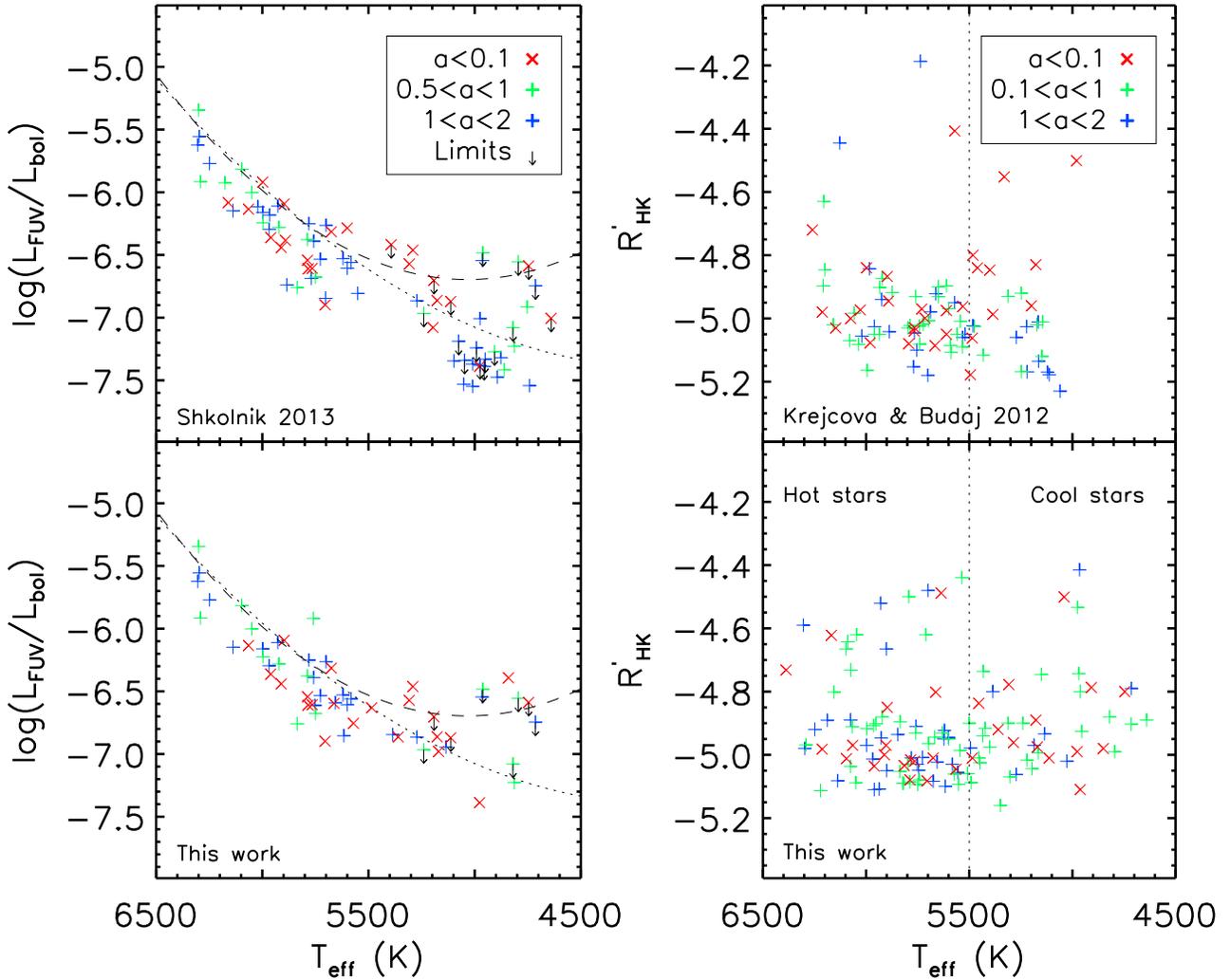} \figcaption{\small UV and
  \ion{Ca}{2} H and K indicators of activity for previously published
  (top) and our FGK MS (bottom) samples as a function of stellar
  effective temperature. Values of $L_{\rm FUV}/L_{\rm bol}$ are from
  Shkolnik (2013) and values of $R_{\rm HK}^{'}$ are from Krej{\v
    c}ov{\'a} \& Budaj (2012; top) or the EOD (bottom). Systems with
  close-in, intermediate, or distant planets, defined as in the
  legend, are colored red, purple, or blue, respectively. The dashed
  curves, left, give the $L_{\rm FUV}/L_{\rm bol}-T_{\rm eff}$
  relation used in Shkolnik (2013); the dotted curves show an ad-hoc
  alternative. The vertical line, right, separates hot from cool
  stars.}
\end{figure*}

\section{UV luminosity and \ion{Ca}{2} H and K activity and post-MS evolution}

Within an inhomegeous sample spanning a wide range in stellar
effective temperature and/or evolutionary stage, it is necessary to
consider intrinsic differences in the types of planets preferentially
hosted by particular stars. Hotter and more massive stars (i.e., FG
versus KM) are more likely to host giant planets (e.g., Fischer \&
Valenti 2005; Gaidos et al.~2013), possibly a result of core accretion
proceeding efficiently in more massive protoplanetary disks (e.g.,
Laughlin et al.~2004). Evolved subgiant stars are less likely to host
hot Jupiters (Johnson et al.~2010), plausibly due to tidal destruction
which can operate efficiently on Gyr timescales out to several tenths
of an AU given the larger stellar radii in subgiants (Schlaufman \&
Winn 2013). Empirically, the distribution of Jovian-mass (0.1-13)
exoplanets within the EOD includes a greater fraction at $a<0.1$ for
stars with $T_{\rm eff}>5500$~K. However, within our FGK MS sample,
$L_{\rm X}/L_{\rm bol}$ shows a similar trend as $L_{\rm X}$ with
$M_{\rm P}/a^{2}$ (the importance of this check is described by
Poppenhaeger et al.~2011), and the extreme systems that drive the
correlation span a wide range of temperatures. In addition, there is
no significant dependence of $L_{\rm X}$ or $L_{\rm X}/L_{\rm bol}$ on
$T_{\rm eff}$ for these stars. By requiring $\log{g}>3.8$ and inferred
stellar radius $R_{*}<2 R{\odot}$ we have effectively excluded
subgiants.

Two additional activity indicators that have been suggested to
correlate with planetary properties are UV luminosity and $R_{\rm
  HK}^{'}$.\footnote{Note that an observed correlation between
  chromospheric stellar activity and hot Jupiter surface gravities
  might indicate that atmospheric mass-loss produces circumstellar
  absorbing material that reduces $R_{\rm HK}^{'}$ (Hartman 2010;
  Fossati et al.~2013; Lanza 2014).} Shkolnik (2013) identify a
marginal increase in relative UV luminosity for close-in systems
($a<0.1$~AU), and Krej{\v c}ov{\'a} \& Budaj~(2012) find evidence that
cooler stars (with $T_{\rm eff}<5500$~K) are more active when they
contain close-in planets, which contrasts with the findings of
Canto-Martins et al.~(2011) of no dependence of $1/a$ on $R^{'}_{\rm
  HK}$. Both Shkolnik (2013) and Krej{\v c}ov{\'a} \& Budaj~(2012)
stress that selection effects could be the underlying cause of the
trends, and we find that this may indeed be the case, albeit related
to stellar evolutionary properties rather than radial velocity
sensitivity bias (Figure~10). We confirm that a positive slope is
present when comparing $L_{\rm FUV}/L_{\rm bol}$ for $a<0.1$ with
$1<a<2$ sytems using the Shkolnik (2013) dataset, but our FGK MS
sample does not contain a significant trend. It appears that their
result is primarily due to a cluster of cool stars that tend to host
distant rather than close-in planets, and that have low UV
luminosities that lie well below the temperature-dependent expected UV
luminosity that is derived from field stars in Shkolnik (2013) and
shown in Figure~10 as a curving dashed line. Either eliminating these
cool stars or employing a closer-to-linear correction, for example the
ad hoc dotted line in Figure~10, would render insignificant the trend
for excess UV luminosity in stars with close-in planets. We also
confirm that KS tests produce a significant difference between
close-in and distant systems for cool stars using the Krej{\v
  c}ov{\'a} \& Budaj~(2012) dataset, but again our FGK MS sample does
not reproduce this result, and the mean and median $R^{'}_{\rm HK}$
values (where measured) for cooler stars are nearly identical and
statistically indistinguishable in systems with $a<0.15$ and $a>0.15$
AU. Here too, it appears that cool and low-activity stars that tend
not to have close-in giant planets help produce the observed trend. As
seen in Figure~10, these type of systems are generally much less
common within our sample.

The specific cool low-activity stars included in these samples have
almost no overlap between the works, but in both cases include
subgiants. For example, the cluster of stars with $T_{\rm eff}<5200$~K
and $L_{\rm UV}/L_{\rm bol}<-7.3$ in Shkolnik (2013) includes eight
objects (HD 5319, 24 Sex, HD 95089, 11 Com, HD 167042, HD 181342, HD
210702, and HD 212771) that are not present in our sample because
their inferred radii are $R_{*}>3 R_{\odot}$ (and 6/8 are additionally
at $d>60$~pc). The SIMBAD luminosity classes of these eight objects
are either III, IV, or unknown. The cluster of stars with $T_{\rm
  eff}<5300$~K and $R_{\rm HK}^{'}<-5.1$ in Krej{\v c}ov{\'a} \&
Budaj~(2012) includes one object in our sample (HD 11964, with a
somewhat hotter $T_{\rm eff}=5350$~K) and six excluded (HD 82886, HD
96063, HD 175541, HD 192699, HD 200964, and again HD 212771) because
their inferred radii are $R_{*}>2 R_{\odot}$ (and 6/6 are additionally
at $d>60$~pc). Many of the cool low-activity stars in Shkolnik (2013)
and Krej{\v c}ov{\'a} \& Budaj~(2012) are described as ``retired A
stars'' by Johnson et al.~(2008, 2010, 2011; see also Lloyd 2011 and
Johnson et al.~2013). Rather than directly supporting magnetic
star-planet interaction, the results of the Shkolnik (2013) and
Krej{\v c}ov{\'a} \& Budaj~(2012) studies may instead reflect the
empirical scarcity of hot Jupiters hosted by subgiants, which effect
might itself arise from cumulative star-planet tidal interaction
producing eventual planetary infall and destruction (Schlaufman \&
Winn 2013).

\section{Conclusions}

We have conducted a comprehensive investigation of the statistical
observability of star-planet interaction, with the following main
results:

1. A sample of 23 solar analogs, including 12 newly observed with {\it
  Chandra\/}, shows no evidence for planet-induced enhancements in
\hbox{X-ray} luminosity. Specifically, the slopes for $M_{\rm P}/a^2$
or $1/a$ versus either $L_{\rm X}$ or $L_{\rm X}/L_{\rm bol}$ (fit in
loglog space) are consistent with zero, and exclude systematic
increases of $\simgt$3 in $L_{\rm X}$ for hot Jupiter systems at
$\simgt$94\% confidence.

2. A sample of 198 FGK main-sequence planet-hosting stars (including
the 23 solar analogs), of which 62 are X-ray detected, does display a
significant correlation between X-ray emission and planetary
properties. While selection biases are present, this trend is
primarily driven by a handful of extreme systems which are here X-ray
luminous.

3. However, the X-ray lumimosities of these extreme systems are
consistent with their chromospheric activity, in contrast to published
scenarios for magnetic star-planet interaction which predict
relatively greater X-ray increases. After removing the $L_{\rm
  X}-R_{\rm HK}^{'}$ relation, the full sample no longer shows a
significant correlation between X-ray emission and planetary
properties.

4. We postulate that the apparently genuine paucity of inactive FGK MS
systems hosting hot Jupiters may result from cumulative tidal
interactions, such as planetary spin-up of the host star plausibly
followed by planetary infall, destruction, and stellar spin-down on
Gyr timescales.

5. There is no significant difference in either UV luminosity or
$R_{\rm HK}^{'}$ for hot Jupiter versus other planetary systems in our
FGK MS sample. This contrasts with some published results, which we
demonstrate are strongly influenced by low-activity cool stars that
are likely post-MS subgiants that either never formed or destroyed
their hot Jupiters.

In summary, we find no positive statistical evidence for magnetic
star-planet interaction acting to enhance coronal activity in hot
Jupiter systems. However, the cumulative tidal influence of close-in
gas giants on their host stars, and perhaps destructively vice-versa,
may explain why the most extreme systems generally do have active
hosts.

\acknowledgments 

We thank Jan Budaj for helpful conversations, an anonymous referee for
constructive comments, and Saul Rappaport for sharing his {\it
  Chandra\/} observation of Arp 143 which serendipitously covers
XO-5. We gratefully acknowledge support for this work from Chandra
Award Number 13200853. This research has made use of the SIMBAD
database, operated at CDS, Strasbourg, France. This research has made
use of the Exoplanet Orbit Database and the Exoplanet Data Explorer at
{\tt exoplanets.org}. The Center for Exoplanets and Habitable Worlds
is supported by the Pennsylvania State University, the Eberly College
of Science, and the Pennsylvania Space Grant Consortium.

\end{document}